\documentclass[12pt,preprint]{aastex}
\def\msun{\rm M_{\sun}}

\def\av{${\rm A_V}$}

\begin{document}
\shortauthors{Hernandez et al.}
\shorttitle{ {\em Spitzer} in $\gamma$ Velorum cluster}
\title{A Spitzer view of protoplanetary disks in the $\gamma$ Velorum cluster}

\author{Jes\'{u}s Hern\'andez\altaffilmark{1,2}, Lee Hartmann\altaffilmark{1}, Nuria Calvet\altaffilmark{1}, 
R. D. Jeffries \altaffilmark{3}, R. Gutermuth \altaffilmark{4}, J. Muzerolle\altaffilmark{5}, J. Stauffer\altaffilmark{6}}

\altaffiltext{1}{Department of Astronomy, University of Michigan, 830 Dennison Building, 500 Church Street, Ann Arbor, MI 48109, US}
\altaffiltext{2}{Centro de Investigaciones de Astronom\'{\i}a, Apdo. Postal 264, M\'{e}rida 5101-A, Venezuela.}
\altaffiltext{3}{Astrophysics Group, School of Physical and Geographical Sciences, Keele University, Keele, Staffordshire ST5 5BG}
\altaffiltext{4}{Harvard-Smithsonian Center for Astrophysics, 60 Cambridge, MA 02138, US}
\altaffiltext{5}{Steward Observatory, University of Arizona, 933 North Cherry Avenue, Tucson, AZ 85721, US}
\altaffiltext{6}{{\em Spitzer} Science Center, Caltech M/S 220-6, 1200 East California Boulevard, Pasadena, CA 91125}

\email{hernandj@umich.edu}

\begin{abstract}
We present new {\em Spitzer Space Telescope} observations of stars in
the young ($\sim$5 Myr) $\gamma$ Velorum stellar cluster.  Combining optical and 2MASS photometry, 
we have selected 579 stars  as candidate members of the cluster.  With the
addition of the Spitzer mid-infrared data,
we have identified 5 debris disks around A-type stars, and 5-6 debris disks 
around solar-type stars, indicating that the strong radiation field in the 
cluster does not completely suppress the production  of planetesimals in the 
disks of cluster members. However, we find some evidence that the frequency 
of circumstellar primordial disks is lower, and the IR flux excesses are smaller than for
disks around stellar populations with similar ages.   This could be
evidence for a relatively fast dissipation of circumstellar dust by the
strong radiation field from the highest mass star(s) in the cluster. Another 
possibility is that  $\gamma$ Velorum stellar cluster is slightly older than 
reported ages and the the low frequency of primordial disks reflects the fast 
disk dissipation observed at $\sim$5 Myr. 

\end{abstract}


\section{Introduction}
\label{sec:int}

In recent years the sensitivity of the {\em Spitzer Space Telescope} has made it
possible to expand our knowledge of dusty disk emission dramatically, enabling
statistically significant studies of disk emission as a function of mass and age
over a much wider wavelength range than previously possible
\citep[e.g.,][]{lada06,aurora06,hernandez07a,hernandez07b}.
In addition to confirming the rapid decrease of optically-thick disk emission previously
inferred from near-infrared observations \citep{strom89, haisch01},
{\em Spitzer} observations have provided strong evidence that disk dissipation timescales
depend significantly on host star mass
\citep{lada06,carpenter06,hernandez07a}, with a timescale for inner disk dissipation of 5-7 Myr 
\citep{haisch01,hartmann05b,hernandez07a} for low mass stars ($\lesssim$1M$\sun$), and $<$3 Myr
\citep{hernandez05} for intermediate mass stars (2-10{$\msun$}).
{\em Spitzer} observations have also provided new clues to disk structure through
Spectral Energy Distribution (SED) analysis, from slight indications of faster
dust settling in inner disks \citep{aurora06} to strong evacuation of the inner
disks in some stars that retain optically thick outer disks \citep{calvet02,calvet05,dalessio05a,espaillat07a}.
Spitzer data have also been used to show that debris disks 
around intermediate mass stars are more frequent 
and have larger 24 $\mu$m excesses around 10-13 Myr,
when models predict that dust is produced by collisional 
cascades due to the formation of large solid bodies ($\sim$1000 km)
in the disk \citep{kenyon05,hernandez06,currie08}.

The possible effects of environment on disk frequency and structure are
less clear.  In clusters with high mass stars,
disk photoevaporation due to the strong ultraviolet radiation 
field can be important, and is directly observed in the Orion Nebula Cluster
\citep{richling00,adams04,hollenbach04,throop05,clarke07,megeath07}. 
\citet{balog07} found a reduction in disk frequency of about a factor of two
in the relatively young cluster NGC 2244, but only in the innermost regions 
(d$\lesssim$0.5 pc).  NGC 2244 is very rich and
contains seven O type stars, so that the ultraviolet radiation
field should be very strong.

The discovery of a young cluster of low-mass stars around the $\gamma$ Vel
system \citep{pozzo00}, with the central binary consisting of an O7.5 star and 
a Wolf Rayet star \citep{demarco99,demarco00},
provides an opportunity to study the possible effects of central star ultraviolet
radiation fields on disk evolution.  
The Wolf Rayet star is classified as WC8 (carbon-rich),
implying a progenitor mass of $>$40$\msun$ \citep{demarco99,crowther07}. 
In this paper we report {\em Spitzer}
observations of infrared emission in this cluster.
In \S\ref{sec:obs} we present the observational 
data used in this contribution, and discuss membership in 
\S\ref{s:mem}. We analyze the infrared data 
\S\ref{sec:disks}, and compare with disk frequencies in other young stellar populations
in \S\ref{sec:comp}, with our main results summarized in \S\ref{sec:conc}.

\section{Observations}
\label{sec:obs}

\subsection{Infrared photometry}
\label{sec2:ir}

We have obtained near-infrared (NIR) and mid-infrared photometry  
of the $\gamma$ Velorum cluster using the four channels 
(3.6, 4.5, 5.8 \& 8.0 {\micron}) of the InfraRed Array Camera 
\citep[IRAC, ][]{fazio04} and the 24 {\micron} band  of the Multiband 
Imaging Spectrometer for Spitzer \citep[MIPS;][]{rieke04}  
on board the Spitzer Space Telescope. These data were collected during 
2005 February 23 (IRAC) and April 10 (MIPS) as part of a GTO program
conducted by the IRAC and MIPS instrument teams.

The IRAC observations were done using a standard raster map with 280" offsets,
to provide maximum areal coverage while still allowing $\sim$20'' overlap
between frames in order to facilitate accurate mosaicking of the data. 
The mapped region was covered by a 9x10 position mosaic, 
with three dither exposures at each position. Images were obtained in high 
dynamic range (HDR) mode, in which a short integration (1 s) is 
immediately followed by a long integration (26.8 s). 
Standard Basic Calibrated Data (BCD) products from version S14.0.0 of 
the {\em Spitzer} Science Center's IRAC pipeline were used to make 
the final mosaics. Post-BCD data treatment was performed using 
custom IDL software \citep{gutermuth04}
that includes modules for detection and correction of bright source 
artifacts, detection and removal of cosmic ray hits, construction 
of the long and short exposure HDR mosaics, and the merger of those 
mosaics to yield the final science images. 
The final mosaics have a scale of 1".22 pixel$^{-1}$.
Point source detections 
were carried out individually on each IRAC band using PhotVis (version 1.09),
an IDL GUI-based photometry visualization tool developed by R. Gutermuth using the DAOPHOT modules
ported to IDL as part of the IDL Astronomy User Library \citep{landsman93}. 

More than 40,000 sources were detected in at least one IRAC band.
We extracted the photometry of these objects using the {\it apphot} package in IRAF,
with an aperture radius of 3\arcsec.7 and a background annulus from 3\arcsec.7 to 8\arcsec.6.
We adopted zero-point magnitudes for the standard aperture radius (12\arcsec) and
background annulus (12-22\arcsec.4) of 19.665, 18.928, 16.847 and 17.391 in the
[3.6], [4.5], [5.8] and [8.0] bands, respectively. Aperture corrections were made using 
the values described in IRAC Data Handbook \citep{reach06}. Final photometric errors 
include the uncertainties in the zero-point magnitudes ($\sim$0.03 mag).

MIPS observations were made using medium scan mode with full-array
cross-scan overlap, resulting in a total effective exposure time per pointing
of 40 seconds.  The images were processed using the MIPS instrument team Data 
Analysis Tool (DAT), which calibrates the data and applies a distortion correction 
to each individual exposure before combining it into a final mosaic \citep{gordon05}. 
A second flat correction was also done to each image using a median
of all the images in order to correct for dark latents and scattered light
background gradients.
The second flat correction is determined by creating a median image of all the data
in a given scan leg, and then dividing each individual image in that scan leg by that median.
Bright sources and extended regions are masked out of the data before creation of
the median, so the nebulosity seen in the map has a negligible effect on any noise
that might be added by the second flat.  We obtained point source photometry 
at 24 {\micron} with IRAF/{\it daophot} point spread function fitting, 
using an aperture size of about 5\arcsec.7 and an aperture correction 
factor of 1.73 derived from the STinyTim PSF model. The aperture size 
corresponds to the location of the first airy dark ring, and was chosen 
to minimize effects of source crowding and background inhomogeneities. 
The sources were identified using an IDL photometry routine, based on 
the IRAF/{\it daophot} package, with a 10$\sigma$ cutoff.  Any obvious 
point sources missed by the find routine were added in by hand. 
The absolute flux calibration uncertainty is $\sim$4\% \citep{engelbracht07}.
The photometric uncertainties are dominated by the background/photon noise.
Our final flux measurements are complete down to about 0.8mJy.

\subsection{Optical photometry}
\label{sec2:opt}

CCD photometry was obtained with the 0.9m Telescope 
at the Cerro Tololo Interamerican  Observatory (CTIO) 
using a Tek 2048x2048 CCD (13.5x13.5 arcmin$^2$). 
A region of 0.93x0.95 deg$^2$ centered on  the $\gamma$ Velorum system was 
surveyed in BVI$_c$ with short (20, 10, 6s) and long (300, 120, 60 s) 
exposure times \citep[more details in][ R. D. Jeffries et al, in preparation]{pozzo00}.
Photometry was performed on the reduced data using the
optimal extraction algorithm detailed in 
\citet{naylor98} and \citet{naylor02}. 
Stars closer than 3{\arcmin} to the central object have poor 
photometry because of the brightness of the $\gamma$ Velorum system.
The optical catalog includes
24,730 sources in the IRAC field ranging from V=10 to V=21 and with 
photometric errors less than 0.1 magnitudes. We augmented the V,B optical 
data set toward the brightest objects using the \citet{kharchenko01} catalog.
A preliminary list of 15,712 objects (hereafter Sample 1) was created
by selecting the optical sources that have photometric measurements 
in all IRAC bands within 3{\arcsec} of matching radius. We have cross-correlated 
Sample 1 with stars in the 2MASS point source catalog. 
About 30\% of these sources do not have 2MASS counterparts within a 
3{\arcsec} matching radius; however, many of these objects without a 2MASS counterpart 
are likely to be background sources (see \S \ref{s:mem}). 
In general, optical, 2MASS, and IRAC positions agree within 1{\arcsec}.

Sample 1 was cross-correlated with the Second XMM-Newton Serendipitous 
Source Catalog (2XMM \footnote{http://xmmssc-www.star.le.ac.uk/Catalogue/2XMM/})
using a correlation radius of 3\arcsec, which is the maximum astrometric 
error expected for most of the sources in 2XMM. 
In an area of $\sim$0.25 square degrees around the $\gamma$ 
Velorum system about 174 objects in Sample 1 have X ray counterparts.

Finally, 32 stars were confirmed as members of the $\gamma$ Velorum 
cluster using the presence of the \ion{Li}{1} 6707 line  in absorption, 
the equivalent width of the H$\alpha$ line and/or radial velocity 
in a narrow range around 17 km/s \citep[][; Jeffries R.D. et al in preparation]{jeffries00}.
The spectroscopic data were obtained at CTIO using the 4-m Blanco telescope in
conjunction with the Hydra fiber spectrograph (resolving power $\sim$ 25000).

\begin{figure}
\epsscale{0.99}
\plotone{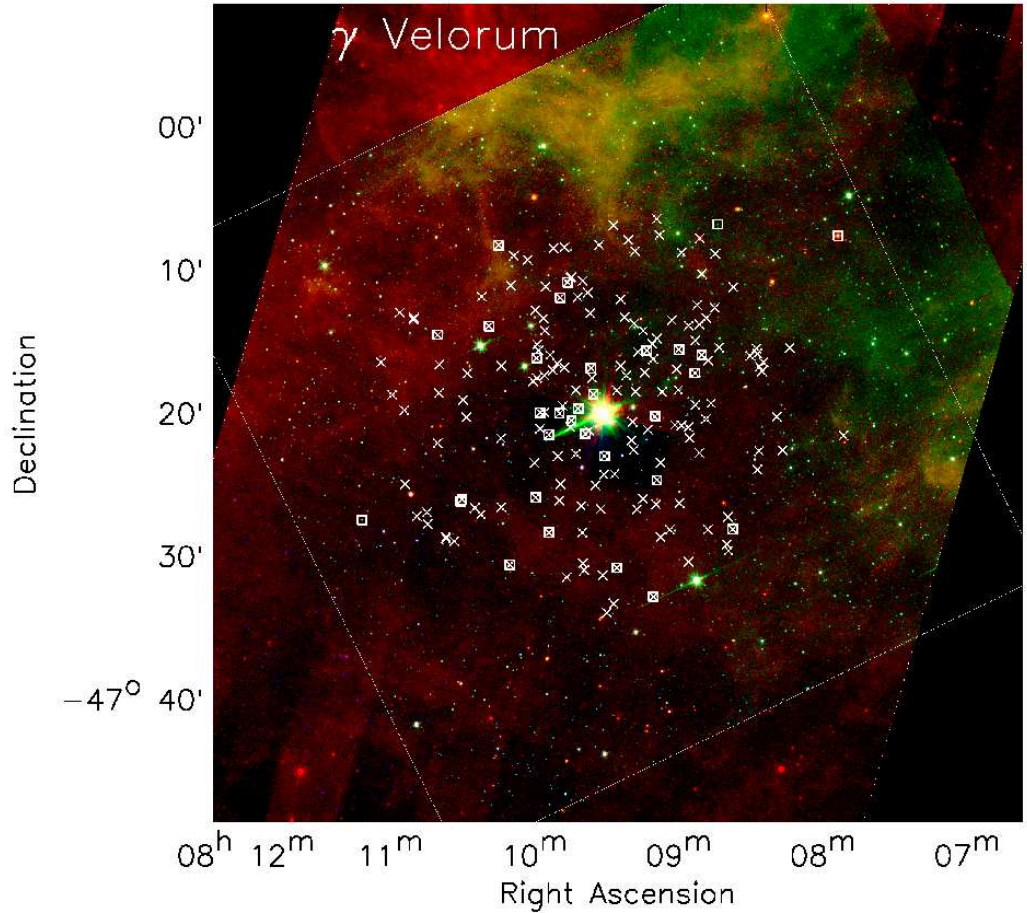}
\caption{ False color image of the $\gamma$ Velorum cluster. It 
is a three-color composite of IRAC images, 3.6 {\micron}(blue) and 4.5 {\micron}(green), 
and MIPS image, 24 \micron(red). The plot is centered on the $\gamma$ Velorum 
binary system. Solid and dashed lines show the IRAC and MIPS fields, respectively.
Members confirmed spectroscopically by Jeffries R.D. et al (in preparation) are plotted as open squares. 
Crosses indicate the location of the X-ray sources from 2XMM catalog.}
\label{fig:map}
\end{figure}

Figure \ref{fig:map} shows the field covered by the optical survey of the 
$\gamma$ Velorum cluster. Solid lines indicate the region studied in 
this paper (IRAC field), which is the area covered by the four IRAC 
channels and where Sample 1 is located. The MIPS observation (dashed lines) 
covers most ($\sim$90\%) of this area. The 2XMM catalog covers about 30\% 
of the IRAC field; X ray sources are indicated on the image (crosses). Most 
of the members confirmed spectroscopically (open squares) are X ray sources.
Only three spectroscopic members do not have X-ray counterparts, and they
are all located outside the 2XMM field of view.
The ring-like diffuse emission structure around the $\gamma$ 
Velorum cluster confirms that some of the gas from which
the cluster formed is still present.

\section{Membership selection}
\label{s:mem}

\begin{figure}
\epsscale{1.0}
\plotone{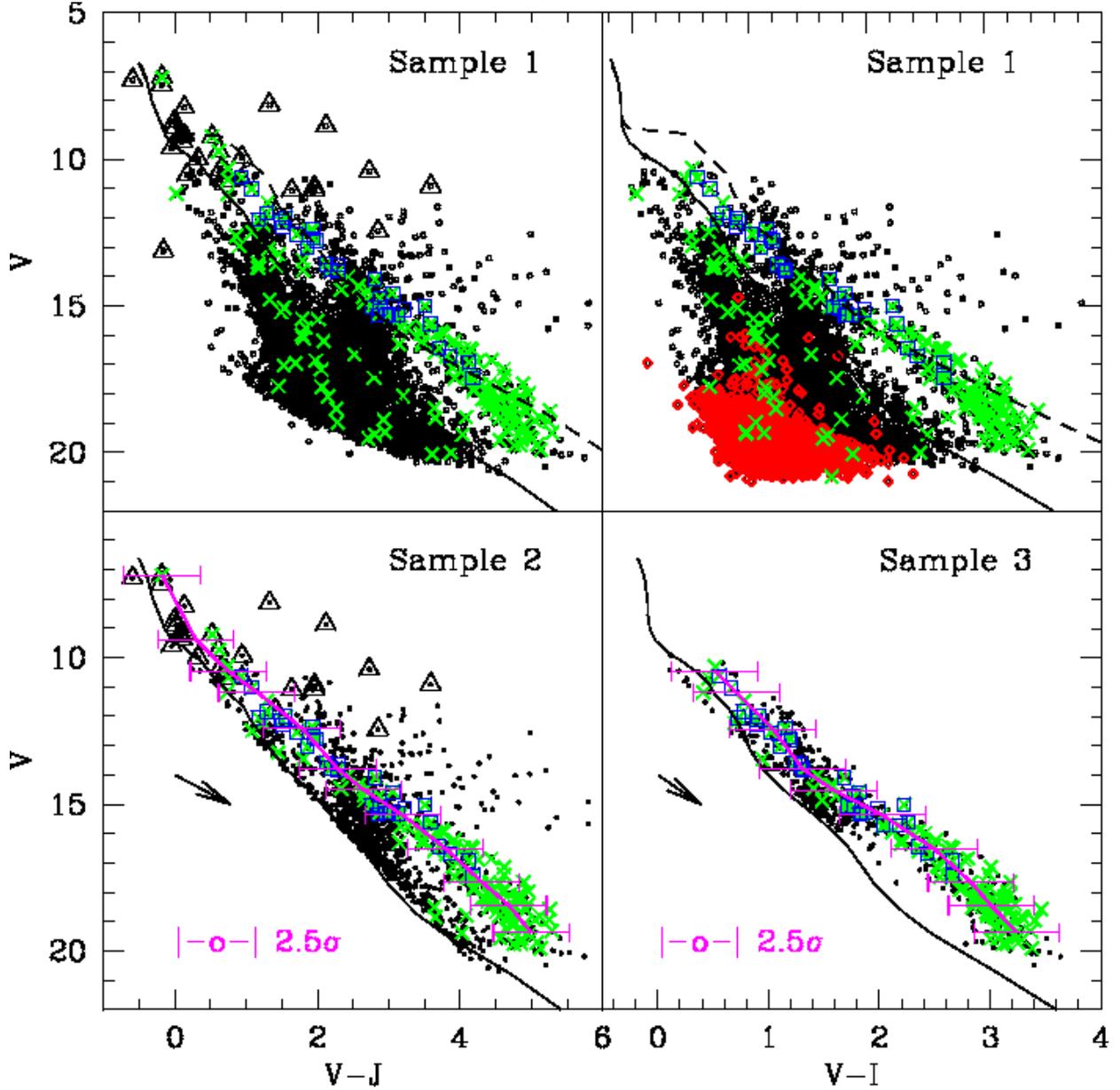}
\caption{ Color magnitude diagrams illustrating the selection of photometric candidates 
of the $\gamma$ Velorum cluster. Open squares and crosses represent the spectroscopic 
members and the X ray sources, respectively. 
In the left panels, open triangles are stars used to augment the optical
data set for the brightest objects \citep{kharchenko01}. In the upper-right panel, 
open diamonds represent stars without 2MASS counterparts, which are likely background objects.
Arrows represent  reddening vectors of \av=1.
In the lower panels, big error bars represent the 2.5$\sigma$ limits from the median colors (thick solid line)  
of the high probability member sample; these error bars define the regions used to select the photometric 
candidates. Assuming a distance of 350 pc, the ZAMS (solid line) and the 5 Myr isochrone 
(dashed line) from \citep{sf00} are plotted in the upper panels 
as reference and do not affect the selection of photometric candidates. 
 }
\label{fig:cmd1}
\end{figure}

Our membership selection is based on using the sequence defined by 
the spectroscopic members \citep[][; Jeffries R.D. et al, in preparation]{jeffries00}
and the X-Ray sources from 2XMM. 
As most of the stars in Sample 1 do not have spectroscopic data, 
we use photometric criteria to make a selection of potential
members based on their colors and magnitudes. Figure \ref{fig:cmd1}
shows color magnitude diagrams (CMDs), V versus V-J (left panels) 
and V versus V-I (right panels), illustrating the procedure 
to select photometric candidates of the $\gamma$ Velorum cluster. 
We assume a distance of 350 pc (see \S \ref{sec:comp}) to plot the Zero Age Main Sequence 
(solid line) and the 5 Myr isochrone (dashed line) from \citet{sf00}.
Since, it is well known that theoretical, non-birthline isochrones
do not accurately match empirical isochrones for intermediate mass stars \citep{hartmann03} and
that the opacity tables at low temperatures are incomplete \citep{lyra06,baraffe98},
these theoretical isochrones are plotted as reference only and 
are not used for the selection of the photometric candidates 
(however, evidence in favor of this age is detailed in \S\ref{sec:comp}).  
Open squares and crosses represent the 
spectroscopic members and the X ray sources, respectively.
Bright stars with optical photometry from \citet{kharchenko01} 
are plotted as open triangles.

The upper left panel shows that X-ray sources are highly concentrated along the nominal
empirical isochrone, with a scattering of stars below this region in the CMD
probably representing background low-mass stars or other non-members.
All the spectroscopic members within the 2XMM field are X ray sources
(see Figure \ref{fig:map}).  Essentially none of the objects above the nominal
isochrone are X-ray sources, and thus are mostly likely heavily-reddened giants.  In addition,
a number of the bright objects from \citet{kharchenko01} are also far above the nominal
isochrone so we reject them as members as well.

As shown in the upper right panel of Figure \ref{fig:cmd1},
stars without 2MASS counterparts (open diamonds) 
are located well below the nominal isochrone and even below the ZAMS (at 350 pc), 
demonstrating that they are background objects. 

In the lower panels we plotted the regions of probable members (large error bars).
These regions were calculated using the median colors (V-I and V-J)
of the member sample (spectroscopic members plus X ray sources above the ZAMS) 
for 1.0 mag bins in the visual band.
We used the differences between the observed colors and the expected colors
(the median; represented by thick solid lines) to calculate the standard deviation
($\sigma$) of the member sample. The regions  of probable members
are defined using 2.5$\sigma$ limit from the median colors 
\citep[see ][]{hernandez07a, hernandez07b}.
The lower left panel shows the stars above the ZAMS on the V versus V-J 
diagram (sample 2), roughly defined using a straight line; 
while the lower right panel  shows the V versus V-I diagram for 
the stars selected using the region of probable members 
in the V versus V-J diagram (sample 3). 

Our final list of photometric candidates includes 579 stars located 
within the probable member region of the CMDs. 
Using the 5 Myr isochrone \citep{sf00} at 350 pc and a reddening of \av=0.2, 
we estimate that we can detect excess at [5.8] and [8.0] 
with signal to noise larger than 5  for a star with spectral type M5 
or earlier (at [3.5] and [4.5] the corresponding signal to noise 
is larger than 40). For MIPS observations, the completeness limit 
roughly corresponds to spectral type early K (see \S\ref{sec2:ir}).
Table \ref{tab:members} shows the optical, IRAC and MIPS photometry
of the photometric candidates of the $\gamma$ Velorum cluster. Column (1) shows the internal 
running identification number in our sample; column (2) provides the 2MASS object name 
\footnote{electronic table also includes 2MASS photometry}; 
columns (3), (4), and (5) give the optical magnitudes in the bands B, V and Ic, respectively;
column (6) indicates the reference for the optical data;
columns (7), (8), (9) and (10) give the IRAC magnitudes in the bands [3.6], [4.5], [5.8] and [8.0], 
respectively; column (11) gives the Flux at the 24{\micron} MIPS band; column (12) 
indicates whether the star is a 2XMM source;
column (13) shows the disk classification based on the IRAC and MIPS analysis (see \S \ref{sec:disks})

\begin{figure}
\epsscale{1.0}
\plotone{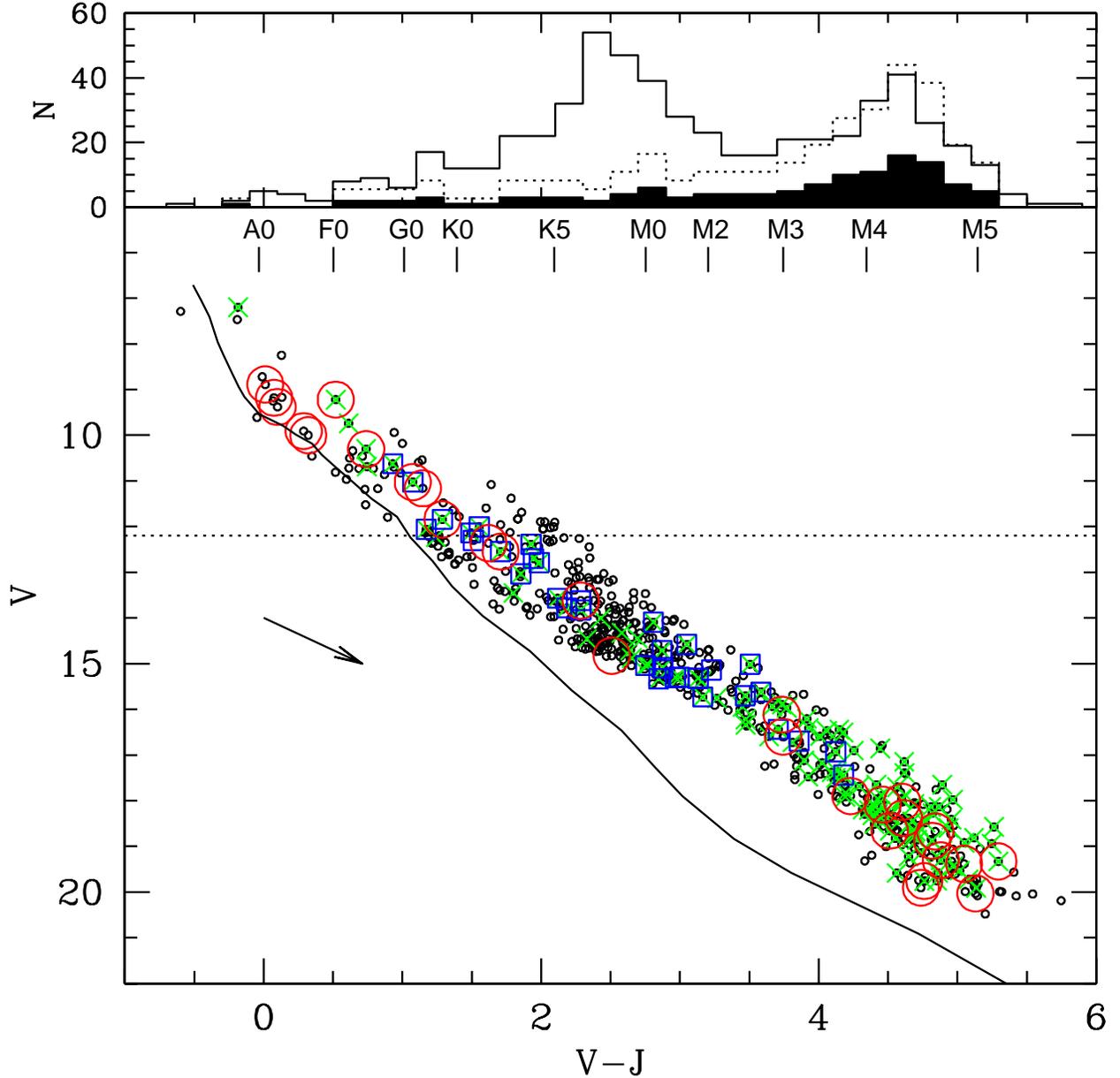}
\caption{ Color magnitude diagram illustrating the contamination level of the photometric candidates sample.
The upper panel shows the V-J distributions for the high probability member sample (solid histogram) 
and the photometric candidates (open histogram).  Using the stars with V-J$>$3.5 
(where the contamination level is low), we normalized both distributions to estimate 
the expected members distribution (dashed histogram).
It is apparent that the number of photometric candidates increases in the color range 
1.5$<$V-J$<$3.5, where the giant branch crosses our sample.   In this range, the 
contamination level by  non-members is quite  high ($\sim$68\%).
In the lower panel, symbols are similar to those in Figure \ref{fig:cmd1}. We plotted the location 
of the spectral type sequence, using the standard V-J color from \citet{kh95}. Stars with infrared 
excess detected in \S \ref{sec:disks} are represented as open circles. The dotted line represents the 
completeness limit for MIPS photometry assuming K-[24]$\sim$0 (photospheric color)  and 
the 5 Myr isochrone \citep{sf00} at 350 pc with a reddening of \av=0.2 (spectral type$\sim$K3).}
\label{fig:cmd2}
\end{figure}

The giant branch of background stars clearly contaminates one portion 
of the cluster member region defined in Figure \ref{fig:cmd1}.
We plotted in Figure \ref{fig:cmd2} the V versus V-J diagram for the 
photometric members illustrating the contamination level expected 
in  several color ranges.  Since the reddening of the $\gamma$ Velorum 
cluster is quite low \citep[\av $<$0.2;][; Jeffries in preparation]{pozzo00}, 
we can estimate spectral types for the sources using their V-J color and 
the standard colors \citep{kh95} indicated in the Figure. The upper panel shows 
the V-J distributions for the photometric candidates (open histogram) 
and for the spectroscopic or X-ray selected member sample (solid histogram). 
Within the 2XMM field, more than $\sim$80\% of the photometric members with
V-J$>$3.5 are X ray sources; most stars without X-ray counterparts are 
very faint (V$>$19.5) or located near the 2XMM field border. 
This suggests that most photometric candidates with V-J$>$3.5  
are real members of the $\gamma$ Velorum cluster. 
Normalizing the member sample histogram to the photometric candidates
histogram in the color range V-J$>$3.5, we estimate the
number of members expected as function of V-J color (dotted histogram).
The deviation between the expected member distribution and the photometric 
candidates distribution indicates the  level of contamination for a given 
range of color (or spectral types).  The number 
of photometric candidates increases between V-J$\sim$1.5 and V-J$\sim$3.5, where the giant 
branch crosses the member sequence. In this range of color, the contamination level by 
non-members of the $\gamma$ Velorum cluster is quite  high ($\ga$68\%).
Since the number of members of stellar groups normally decreases toward higher stellar masses, 
the relative contamination at V-J$<$1.5 can still be significant.  

\begin{figure}
\epsscale{1.0}
\plotone{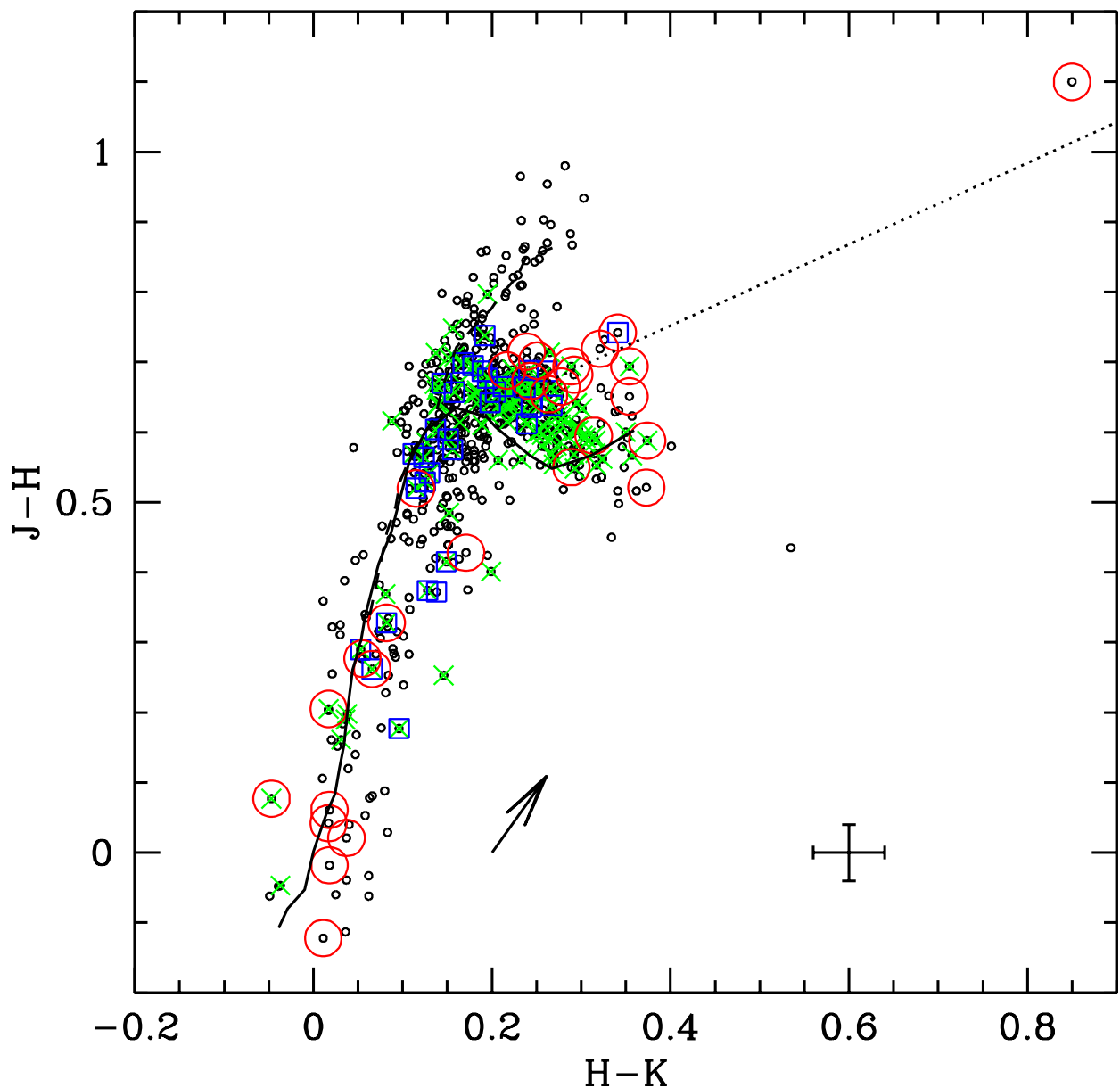}
\caption{ 2MASS color color diagram, J-H versus H-K, illustrating the selection of M giant stars 
in the field of view of the $\gamma$ Velorum cluster. Solid and dashed lines represent the 
standard sequence from \citet{bessell88} for dwarfs and giants, respectively. The arrow indicates
the reddening vector for \av=1. The error bar represents the typical 2MASS color uncertities
in our sample. As a reference, we plot the loci of classical T Tauri stars (CTTS) defined by 
\citet{meyer97}. Other symbols are the same as Figure \ref{fig:cmd2}. }
\label{fig:ccd}
\end{figure}

Figure \ref{fig:ccd} shows the 2MASS color color diagram, J-H versus H-K, for the 
photometric members of the $\gamma$ Velorum cluster. Since the locus of giants 
separates from main sequence late type stars \citep{bessell88},
we can identify background M type giants in this diagram. We have selected 27 stars
in the locus of giants with J-H$\ge$0.8, these stars are labeled in Table 
\ref{tab:members} as M type giant candidates. Taking into account this M giant sample, 
the contamination level  where the giants branch  crosses the $\gamma$ Velorum sequence in Figure
\ref{fig:cmd2} is still high ($\ga$66\%).

\section{Disk diagnostics}
\label{sec:disks}

\begin{figure}
\epsscale{0.8} 
\plotone{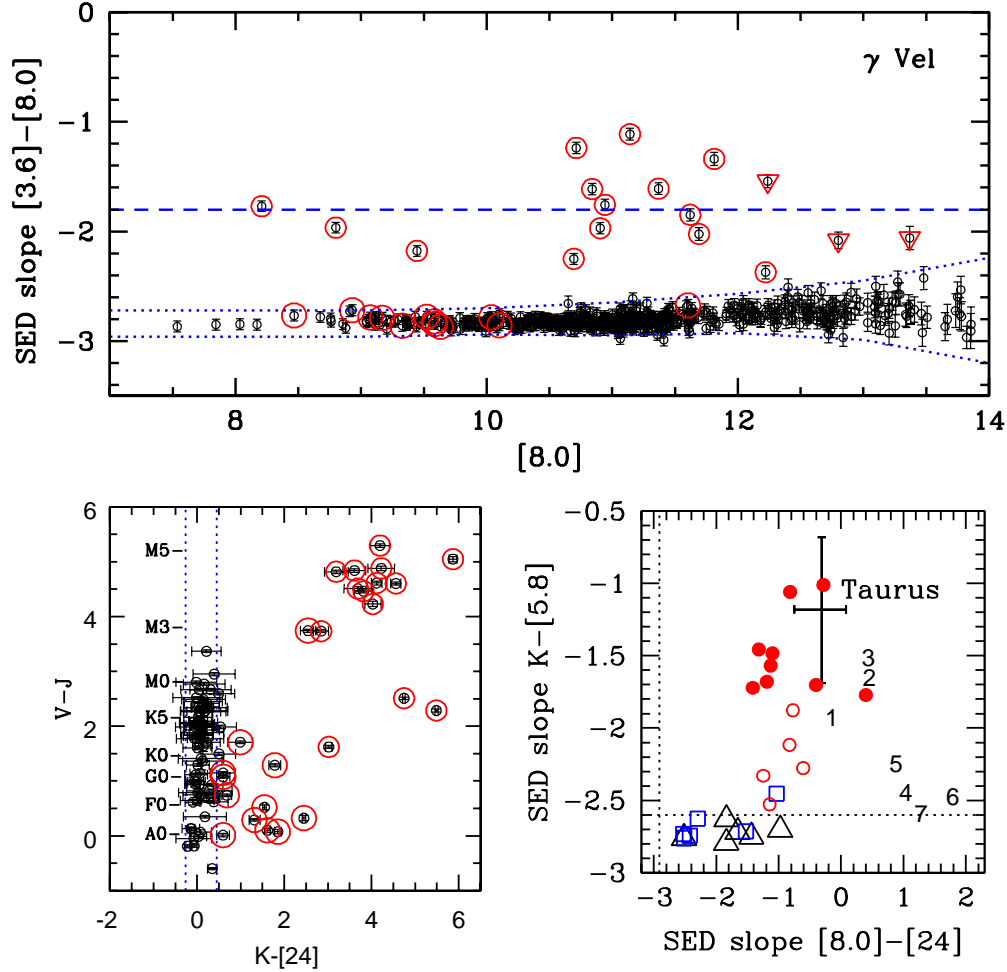}
\caption{ Diagrams illustrating the detection of disks based 
on their infrared emission above the photospheric levels (dotted lines).
The upper panel shows the IRAC SED slope diagram indicating stars with 
excess at 8\micron. The dashed line represents the class II limit 
proposed by \citet{lada06} and the inverse triangles are 
objects with IRAC excess but lacking MIPS detections. Stars with 
24{\micron} excess are represented by points surrounded by large open circles. 
The color magnitude diagram V-J versus K-[24] is displayed in 
the bottom-left panel illustrating the detection of stars 
with excess at 24\micron. The bottom-right panel shows the SED 
slope for K-[5.8] versus the SED slope for [8.0]-[24] diagram. 
Disk bearing stars in different spectral type ranges 
are plotted with different symbols, stars with spectral 
type A are represented with open triangles, 
stars from F  to early-K with open squares  
and late type stars (K5 and later) with circles.
The error bar represents the median and quartiles
for the Taurus disks \citep{hartmann05a,furlan06};
solid circles are stars with disk emission similar 
than Taurus's disk population (class II objects). 
As reference,
we also mark the locations of stars in other 
star-forming regions believed to have  
pre-transitional \citep[1, 2 and 3,][]{espaillat07b,furlan06} and transitional 
disks \citep[4, 5, 6 and 7,][]{calvet02,calvet05,dalessio05b,espaillat07a}. 
One class II object is a pre-transitional disk candidate.}
\label{fig:disk}
\end{figure}

Figure \ref{fig:disk} shows three diagrams used to identify and characterize 
stars bearing disks in the $\gamma$ Velorum cluster. The top panel  shows the 
SED slope, determined from the [3.6]$-$[8.0] color 
(SED slope=$dlog[{\lambda}F_\lambda]/dlog[\lambda]$), 
versus the [8.0] magnitude. Dotted lines represent the 
3 $\sigma$ boundaries (photospheric level), which is calculated using 
the photometric errors propagated from the [3.6]$-$[8.0] color 
\citep{hernandez07a, hernandez07b}. Stars with excess emission at 
8 {\micron} can be identified in this diagram above the photospheric 
limit.  In general, disk bearing 
stars in Taurus \citep{hartmann05a} are located  
above the dotted line, with an IRAC SED slope $>$ -1.8 \citep[see ][]{lada06, hernandez07b}; 
this limit is used to identify objects with optically thick primordial disks (class II). 
Half of the stars with IRAC excesses are located below  the class II region suggesting that the
disks around these stars have evolved significantly, and are  in an intermediate  phase between 
stars with optically thick disks and debris disk stars.
We call these evolved disks.

The bottom left panel shows the V-J versus K$-$[24] color-color diagram,  in which we identify  
stars with 24 {\micron} infrared emission above the photospheric level
\citep[e.g.; ][]{gorlova04,gorlova06, hernandez06,hernandez07a}. The K-[24] color distribution of 
stars with K-[24]$<$1.0 describes a Gaussian centered at  K-[24] = 0.05 with $\sigma$=0.14.
The 3 $\sigma$ boundaries (dotted lines) represent the photospheric region at 24 \micron; 
stars with 24{\micron} excesses are located red-ward from this region.
Except for the three faintest stars with excesses at 8{\micron} (inverse triangles in the upper panel), 
all stars with IRAC excess have excess at 24{\micron}. Our sample of disk-bearing candidates includes
14 stars with excess at 8{\micron} and 24{\micron}, 3 stars with excess at 8{\micron} 
and with no MIPS detection, and 12 stars with excess at 24{\micron} 
and photospheric color in the IRAC bands.

Using the standard V-J colors in Figure \ref{fig:cmd2} (or in the bottom left panel), we can estimate 
the spectral types of stars with infrared excesses.  We use three bins:
early type stars (V-J$\le$0.54; F0 or earlier); solar type stars 
(0.54$<$V-J$<$2.0; from early-F to early-K); and late-type stars (V-J$>$2.0; K5 or later). 
The bottom right panel of Figure \ref{fig:disk} displays the distribution of the disk-bearing stars
in a SED slope diagram, generated using the K-[5.8] and [8.0]-[24] colors. 
The [8.0]-[24] SED slope indicates whether the star has 
a flat or a rising SED (SED slope $\ge$ 0) at wavelengths greater than 8.0{\micron},
and hence whether it should be classified as a
flat-spectrum or a Class I source.

We can also compare the properties of disks in the $\gamma$ Velorum cluster
to those disks with strong evidence of dust evolution.
Primordial disks with a high degree of inner disk clearing
have been called transitional disks; observationally they are characterized by 
relatively small excess  below 5.8{\micron} and a large excess at longer wavelengths
implying a a SED slope [8.0]-[24] $>$ 0 \citep{hernandez07b}.
We show in Figure 4 the location of four well known
transitional disks in Taurus, Chameleon, and the TW Hya association:
TW Hyd (4) , GM Aur (5), Coku Tau (6) and 
CS Cha (7) \citep{calvet02,calvet05,dalessio05b,espaillat07a}.
In addition, we show in Figure 4 the location of the newly
discovered class of pre-transitional disks, in which gaps
in primordial disks rather than holes have been identified
\citep{espaillat07b,brown08}. These are 
LkCa 15 (1), UX Tau (2),  and HK Tau (3) \citep{espaillat07b,furlan06}.  
In contrast to the disk population in the $\sigma$ Orionis cluster, the Orion OB1b association, 
and the 25 Orionis aggregate \citep[ with ages of $\sim$ 3, 5 and 8 Myr, respectively][]{hernandez07a,hernandez07b},
there are no transitional disk candidate detected in the $\gamma$ Velorum cluster.
However, there is a pre-transitional disk candidate (star 52; labeled as PTD in Table \ref{tab:members}) with infrared excesses similar 
to the pre-transitional disk systems UX Tau and Lk Ca 15 \citep{espaillat07b}.

The late-type stars exhibit less infrared excess than the 
disk population in Taurus, suggesting that 
the disks in the $\gamma$ Velorum cluster have evolved substantially.

Except for  one solar type star (see \ref{s:gtype}), 
which exhibits IRAC and MIPS excesses, the early-type 
and solar-type stars show excesses only at 24{\micron}. The relatively 
small 24{\micron} excess (K-[24]$<$3) of these objects are consistent with their
being debris disks, in which second generation dust 
is produced by collisions between planetesimals \citep[e.g.,][]{kenyon05,hernandez06,currie08}.
 However, theoretical models of collisional cascade at 30-150 AU 
\citet{kenyon05} do not predict enough excess emission 
to explain the observed excesses at 5 Myr. A small dust contribution 
from collisional cascade in the inner disk ($<$30 AU) or 
remaining primordial dust in the disk could explain 
these differences \citep[see][]{hernandez06}. Since, the timescale 
for primordial disk dissipation decreases as 
the stellar mass increases \citep[e.g.][]{hernandez07a,lada06}, 
it is more likely to observe remaining primordial dust 
in the disk around solar-type stars than in early-type stars.

Visual inspection of the IRAC  and MIPS images
reveals that stars with infrared excesses are not contaminated by
artifacts, nebulosity or 
extragalactic sources. In general, the stars with infrared 
excesses  are isolated sources. Only stars 115 and 359 
have near faint sources (within the background photometric annulus) 
that could affect the photometry at [3.6] and [4.5]; however,
these objects look like isolated objects at longer wavelengths, 
where infrared excesses are more obvious.  
The star 8797 (with elongated PSF) can be a 
close binary not resolved by IRAC.  For this source, 
IRAC and MIPS photometry includes contributions from both 
components.

\subsection{Disks of early type stars}
\label{s:debris}

\begin{figure}
\epsscale{1.0}
\plotone{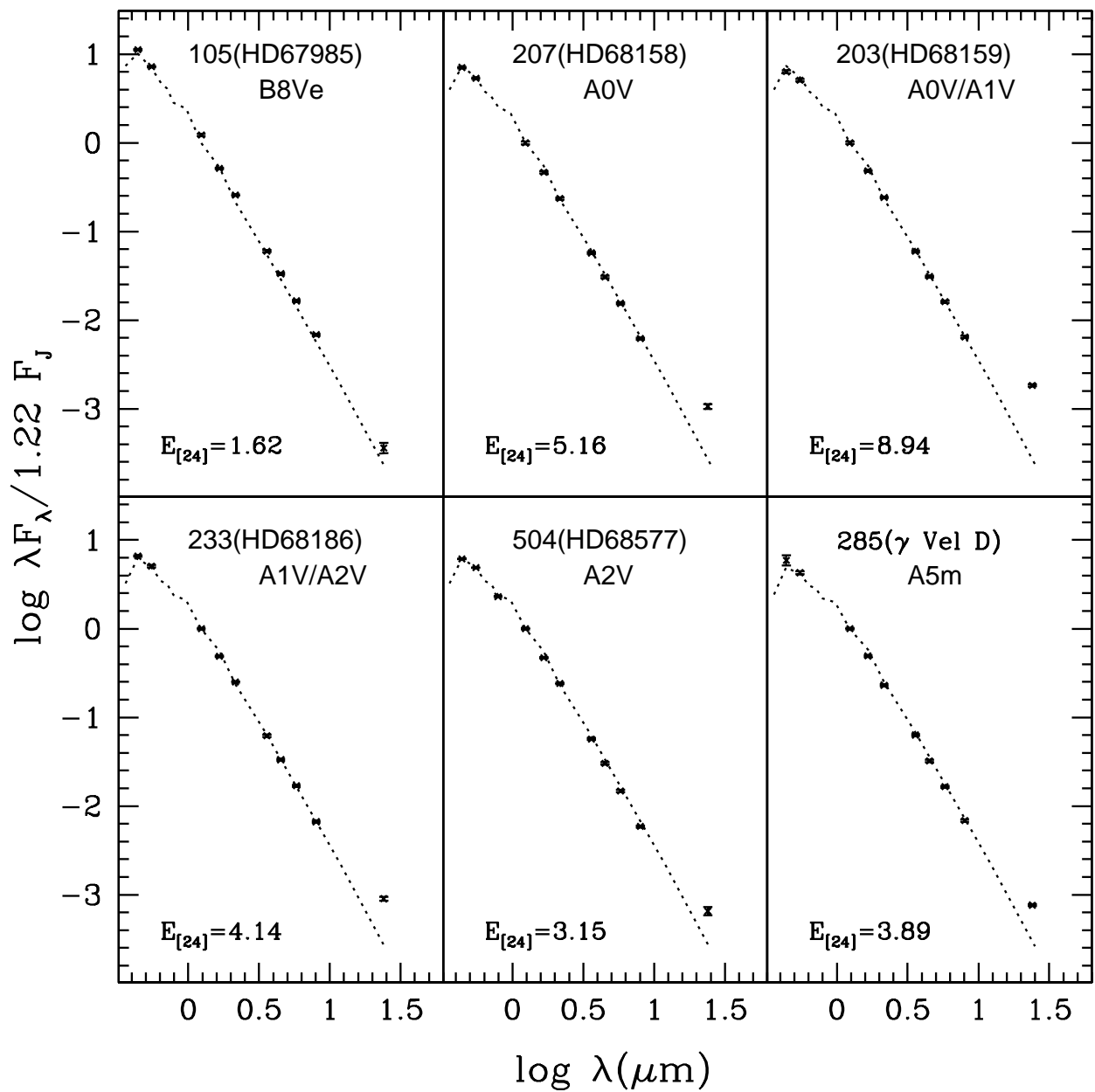}
\caption{ SEDs for A type stars with infrared excess. Each panel 
shows the spectral type of the object \citep{houk78,abt76}, 
and the excess ratio, {\rm$E_{24}$} \citep{rieke05}, calculated from the K-[24] color. 
Dotted lines show the corresponding photospheric colors \citep{kh95}.
Star 105 (upper-left panel) is a Be star, the other five 
stars are debris disk candidates.}
\label{fig:sed1}
\end{figure}

We detected six early type stars with excess at 24{\micron}. Figure 
\ref{fig:sed1} shows the SEDs of these objects. Except for the star 285 
\citep[$\gamma$ Vel D, ][]{abt76}, a binary star near the $\gamma$ Velorum system, 
all stars have spectral types from \citet{houk78}.  \citet{abt76} classified
$\gamma$ Vel D as a metallic-lined binary star with spectral type 
ranging from A2 to A8; we assumed a spectral type of A5 in Figure \ref{fig:sed1}. 

Color excesses (E$_{V - J}$) were obtained by  interpolating spectral types 
in the table of standard V-J color given by \citet{kh95}. 
SEDs were corrected for reddening (\av) using E$_{V - J}$ and 
the extinction relation from \citet{cardelli89}, with the 
value of total-to-selective extinction for normal interstellar reddening ({\rm $R_V = 3.1$}).  
The small values of {\av} obtained (average=0.2$\pm$0.1) support the low 
reddening reported for the $\gamma$ Velorum cluster 
\citep[][Jeffries R.D. et al in preparation]{abt76,pozzo00}. 
We use the color K-[24] to determine the excess ratio at 
24{\micron} \citep[e.g.,][]{rieke05}, {\rm$E_{24}=10^{(K-[24]-0.05)/2.5}$},
where 0.05 is the mean value of K-[24] for the photometric candidates with MIPS detections.  

Star 105 exhibits a very small excess at 24{\micron}. As it is also
an emission-line star \citep{houk78,macconnell81} and its 2MASS colors 
are similar than those observed in classical Be stars \citep{hernandez05}, this excess is likely
to be free-free emission in a hot envelope - that is, this is probably the excess
of a classical Be star.
The remaining stars with 24{\micron} excesses are likely to be debris disk systems 
with {\rm$E_{24}$} similar to those observed 
in debris disks candidates of the $\sigma$ Orionis cluster and the Orion OB1b 
subassociation \citep{hernandez06,hernandez07a}.

Most of stars with V-J$\le$0.54 (F0 or earlier) 
have spectral types from \citet{houk78} or \citet{kharchenko01}. 
Except for the stars 277 and 572, the published spectral type and the 
photometric spectral type calculated from the V-J color (see Figure \ref{fig:cmd2}) 
agrees within 5 spectral subclasses.  However, additional studies (e.g radial velocity) 
are necessary to confirm membership of our early type sample. 
Assuming that stars 277 and 572 are not members of the $\gamma$ Velorum cluster, 
we find 8 early type candidates without infrared excess (diskless), 
5 debris disk candidates and one classical Be star. Since some diskless
could not be members of the cluster, the debris disk frequency 
estimated in this work (5 debris disks / 14 early type stars; $\sim$36$\pm$16\%) 
is a lower limit.

\citet{adams04} report that mechanisms of gas dispersal, like photoevaporation, 
can disrupts planet formation in two important ways: (1) if the gas is dispersed
before the disk dust has grown to centimeter size, the 
formation of planetesimals will be curtailed as the dust is removed with the gas;
(2) if the gas is dispersed before large rocky planets 
are formed, the formation of gas giant planets like Jupiter will be 
suppressed. The debris disk frequency ($\ga$36\%)  and the infrared 
emission from the debris disks in the $\gamma$ Velorum cluster are comparable
to the debris disk population of stellar groups with similar ages.
For instance, using the same method to detect 
stars with excess at 24{\micron}, \citet{hernandez06} reported 
debris disk frequency of $\sim$38\%$\pm$3\% and $\sim$46\%$\pm$4\% 
for the Orion OB1b subassociation ($\sim$5 Myr) and for the 25
Orionis stellar aggregate ($\sim$8 Myr), respectively. 
Since the presence of debris disks implies the formation of km size objects \citep{kenyon05},
proposal (1) appears unlikely.
 
\subsection{Disks around solar-type stars}
\label{s:gtype}

\begin{figure}
\epsscale{1.0}
\plotone{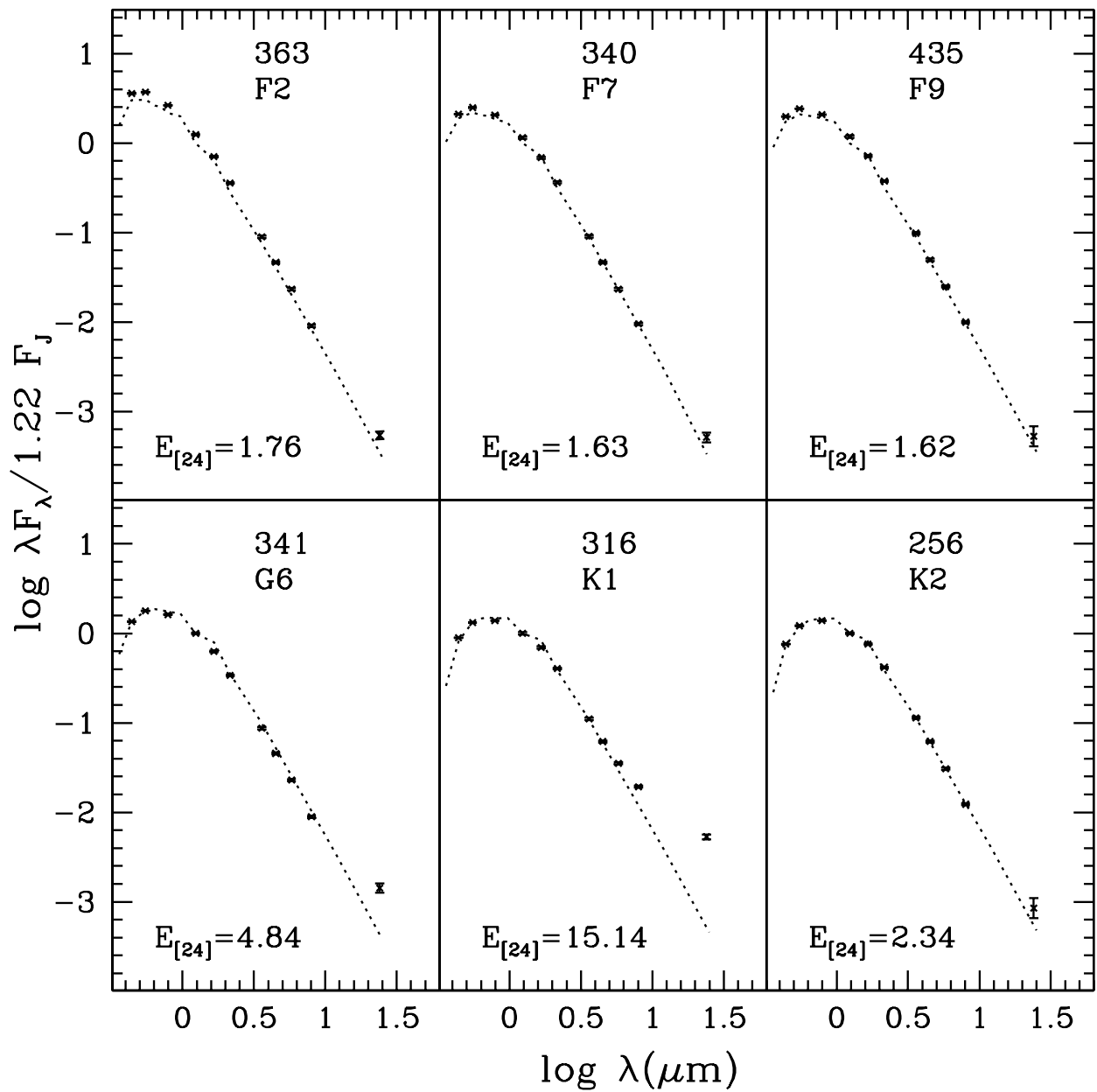}
\caption{ SEDs for solar type stars with infrared excesses.
Each panel shows the photospheric spectral type calculated
from the color  V-J assuming a reddening of \av=0.15 mag and 
intrinsic colors from \citet{kh95}. Dotted lines represent 
the corresponding photospheric level \citep{kh95}. The excess 
ratio at 24{\micron} is displayed in each panel.}
\label{fig:sed2}
\end{figure}

We detected six solar-type stars with excesses at 24{\micron}.
Figure \ref{fig:sed2} shows the SEDs for this subsample. 
Each panel shows the corresponding photospheric fluxes \citep{kh95}.
Spectral types were estimated by interpolating 
dereddened V-J colors using the standard table given by \citet{kh95}.
We assumed a reddening value of \av=0.15 \citep[E$_{V-I}$=0.06;][]{pozzo00}.

The values of {\rm$E_{24}$} in this subsample
are slightly smaller than the debris disks around FGK stars observed  
in the 15 Myr-old Scorpius-Centaurus-Lupus 
sub-association \citep{chen05,siegler07} and comparable to the 
excesses of debris disks around FGK stars in the 30 Myr old cluster NGC 2547 
\citep{gorlova07}. 
Assuming the same excess detection threshold of E$_{24}$=1.41 
(K-[24]=0.42), we have calculated a median E$_{24}$ of
2.1$^{+4.8}_{-1.6}$, 2.8$^{+14.2}_{-2.0}$ and 2.0$^{+2.3}_{-1.5}$
for F G type debris disk candidates in  the $\gamma$ Velorum cluster, 
Scorpius-Centaurus-Lupus sub-association and the NGC 2547 cluster, respectively.
The errors represent the upper and lower quartiles of each sample.
If interpreted as an evolutionary sequence, these
disk emissions would be consistent with a growth period of dust 
production and a following decrease as dust particles
are removed, as predicted by  theory \citep{kenyon05}.  
Similar trends have been seen in A type stars \citep{hernandez06, currie08}.   
However, given the small number of stars, the statistical significance of this
result is low.

Only star 316 shows an excess at shorter wavelengths; it 
represents the earliest star exhibiting excess in the 
IRAC bands. The excesses around this star can be explained 
if the disk is in an intermediate phase evolving from a
primordial to a debris disk (evolved disk). Another possibility is 
that this star has a very massive debris disk produced
recently by a catastrophic collision between two 
planetary-size bodies \citep{gorlova07,kenyon05}.

\subsection{Disks around late type stars}
\label{s:late}
\begin{figure}
\epsscale{1.0}
\plotone{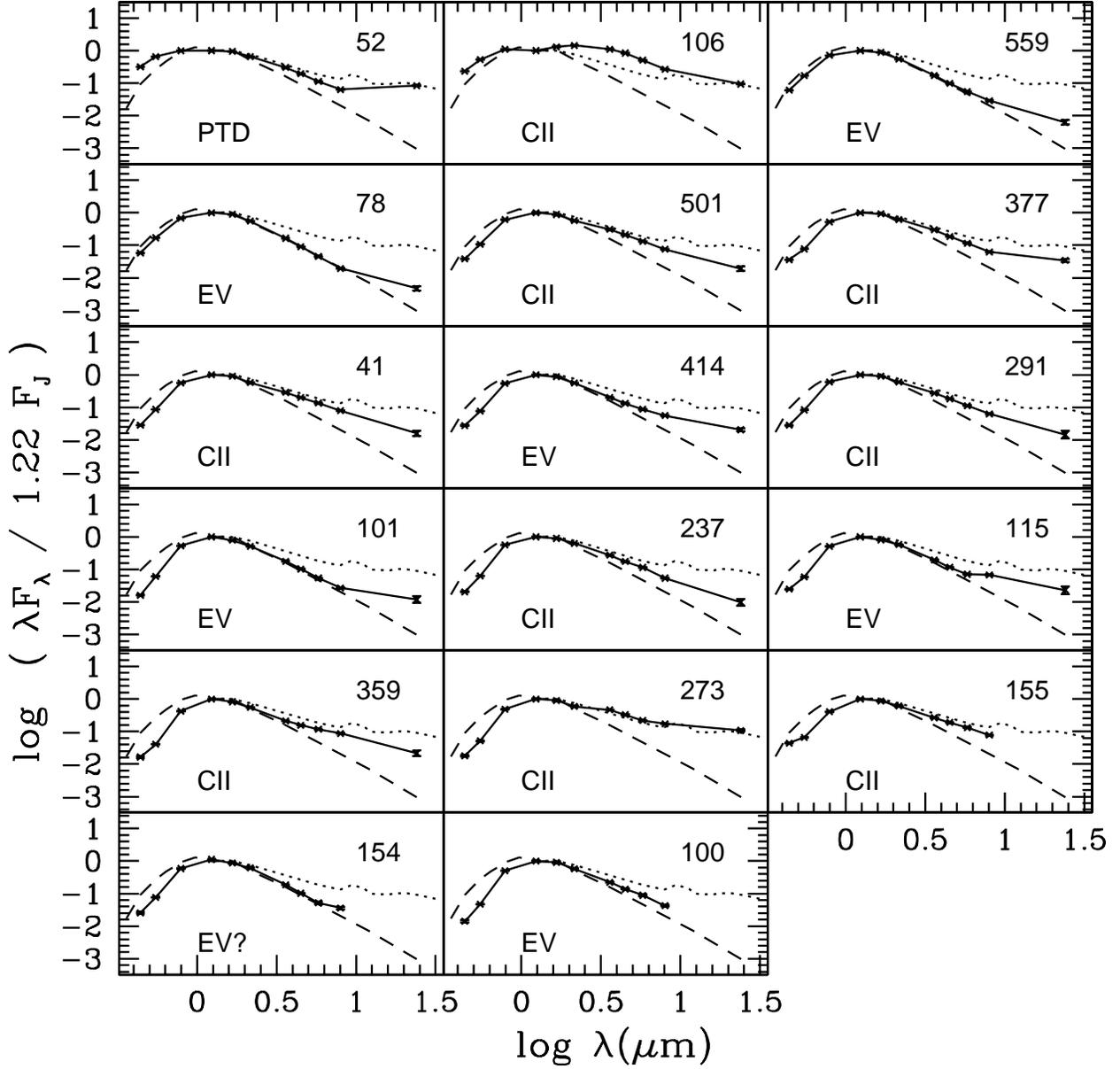}
\caption{ SED for low mass stars with infrared excess (solid lines).
Dotted and dashed lines represent the median SED for 
class II stars in Taurus \citep{furlan06}, and the photospheric 
fluxes for stars with spectral type M2 \citep{kh95}, respectively.
Each panel shows the disk classification: 
pretransitional disks (PTD), class II objects (CII), and 
evolved disk objects (EV).}
\label{fig:sed3}
\end{figure}

Figure \ref{fig:sed3} shows the SEDs of late type stars sorted 
by brightness in the V band. We display the median SED 
of optically thick disk stars in Taurus \citep{furlan06} and 
the photospheric fluxes for a star with spectral type M2 \citep{kh95}.
The diversity of disks in the $\gamma$ Velorum cluster
is apparent. We distinguish between
class II systems (CII, with SED slopes K-[5.8]$>$1.8), 
and evolved disk (EV) systems exhibiting smaller IRAC/MIPS excesses than 
class II systems (K-[5.8]$<$1.8).

Two stars have photometric colors between K5 and M0;
one of them (star 52) shows infrared excesses similar 
to pre-transitional disks objects \citep{espaillat07b}. The
other star (106) exhibits larger infrared excesses than the optically 
thick median in Taurus.
The next two objects (stars 559 and 78), with photometric spectral type $\sim$ M3,
exhibit modest infrared emission, which could be explained by 
decreasing the height of the irradiation surface in the disks due to 
a higher degree of settling producing flatter disk structures 
\citep{hernandez07a, hernandez07b, dalessio06}. 
The remaining 13 stars with infrared excesses have photometric 
spectral types M4 or later (see Figure \ref{fig:cmd2}) 
suggesting, as in other young stellar population \citep{hernandez07a,hernandez07b,carpenter06,lada06} 
that primordial disk frequencies depend on the mass of the central object. 
Using the number of members expected in our photometric sample 
(dashed histogram in Figure \ref{fig:cmd2}), 
we calculated disk fractions of 4.3$\pm$3.0\% for the color range 2$<$V-J$<$3 
($\sim$K5-M1), 3.5$\pm$2.4\% for the color range 3$<$V-J$<$4 ($\sim$M1-M3.5) and
7.1$\pm$2.0\% for stars in with V-J$>$4 (later than M3.5). 

Stars 100, 154, 155 are too faint to be detected by MIPS. Two of them show 
excesses in all IRAC bands, making identification of these stars (100 and 155) as
having dust disks fairly certain.  However, star 154 shows excess only at 8{\micron}, 
which could be produced by either disk emission or PAH contamination. 

\begin{figure}
\epsscale{1.0}
\plotone{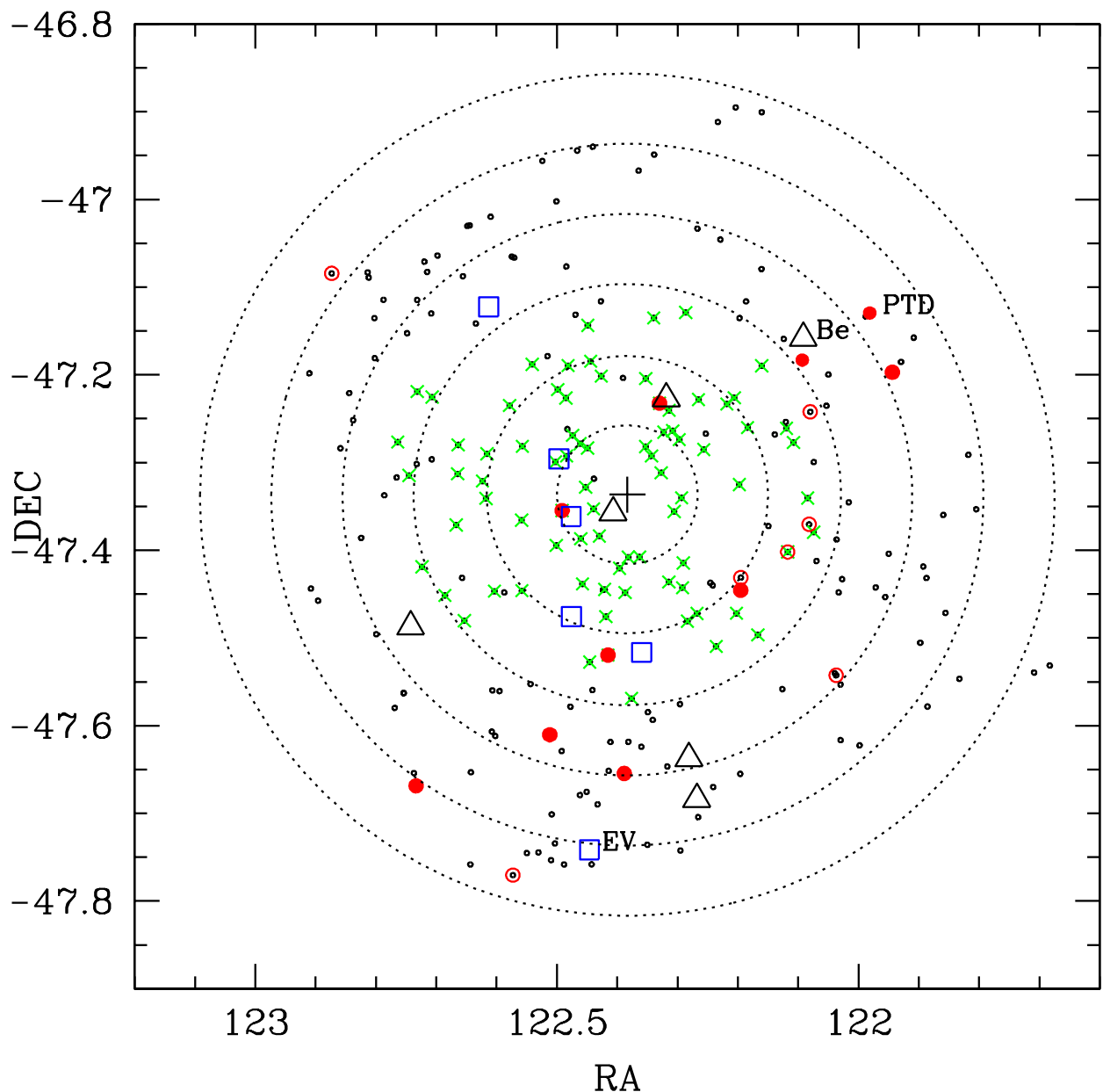}
\caption{Space distribution of stars with infrared excesses. 
Symbols for stars with infrared excess are the same as the bottom right panel of Figure \ref{fig:disk}.
We display X-ray sources (crosses) and  photometric candidates (small open circles) 
with V-J$>$3.5 (little contamination is expected in this color range)
Dotted circles represent projected distance from the $\gamma$ Velorum system
in intervals of 0.5 pc. We mark the position of the solar-type star with 
IRAC excess (EV), the Be star (Be) and the pre-transitional disk candidate(PTD).}
\label{fig:space}
\end{figure}

Figure \ref{fig:space} shows the spatial distribution of the stars 
with infrared excesses in our sample. Different spectral type ranges
are represented with different symbols. As reference, 
we identify X-ray sources and  photometric candidates 
with V-J$>$3.5 (little contamination by non-members is expected 
in this color range).  The projected distances from the $\gamma$
Velorum system are displayed in intervals of 0.5 pc.  One debris disk
is located inside 0.25 pc,
while the only optically-thick 
disks are in systems located at projected distances of $\ga$0.5 pc.
Since, about 10\% of the stars bearing disks are expected to 
have transitional disks \citep[][; Muzerolle J., in preparation]{hernandez07b},
we can expect one or two transitional disks in the $\gamma$ Velorum cluster.
However, we did not  detect any transitional  disk objects in the cluster.
We can speculate that external mechanisms of disk dispersal
(like photodissociation and/or photoionization by strong stellar winds and/or strong UV radiation),
affecting mainly the disk surface and the outer region of the disk,
combined with the fast inner clearing characterizing  transitional 
disks, decrease the probability of observing a disk in this phase. 
Given the small number of stars bearing disks, the statistical 
significance of this speculation is very low.

\section{Comparison with other young stellar populations}
\label{sec:comp}

To compare the disk frequencies in $\gamma$ Vel with other regions, 
we need to estimate the cluster age, which depends to some extent upon
the assumed distance to the cluster.
\citet{burningham06} confirmed that this group is a stellar cluster
belonging to the Vela OB2 association, analogous to the 
$\sigma$ Orionis cluster in the Orion OB1b subassociation \citep{hernandez07a}.
The primary star in the $\gamma$ Vel system is the only WR star 
with a distance from 
Hipparcos \citep[258$^{+41}_{-31}$pc;][]{schaerer97}; but this distance is 
in marked contrast to distances estimated by other methods (350-450 pc) 
and to the Hipparcos distance of the Vela OB2 association 
\citep[410$\pm$12 pc;][]{dezeeuw99}.  Using interferometric 
observations to estimate the geometrical parameters 
of the orbit of the $\gamma$ Vel system, \citet{millour07} 
and \citet{north07} give distances of 368$^{+38}_{-13}pc$ 
and 336$^{+8}_{-7}pc$, respectively. These values are in agreement
with the distance obtained by \citet[][ 349$^{+44}_{-35}$ pc]{leeuwen07} with the 
new reduction of the Hipparcos data that includes modeling of the satellite dynamics 
and eliminates the data correlation in the original catalog caused by 
attitude-modeling errors \citep{maiz08}.
The age estimated 
by \citet[3.5 Myr;][]{north07} for the O star is in 
agreement with the age estimated by \citet[$\lesssim$5 Myr;][]{pozzo00}
for the cluster and the age estimated by \citet[3.6 Myr;][]{demarco99} 
for the $\gamma$ Vel system. 
We assume the corrected Hipparcos distance of 350 pc, which is 
in agreement within 1.5$\sigma$ from the distance calculated by \citet{millour07} 
and \citet{north07} and a reference age of 5 Myr.

\begin{figure}
\epsscale{1.0}
\plotone{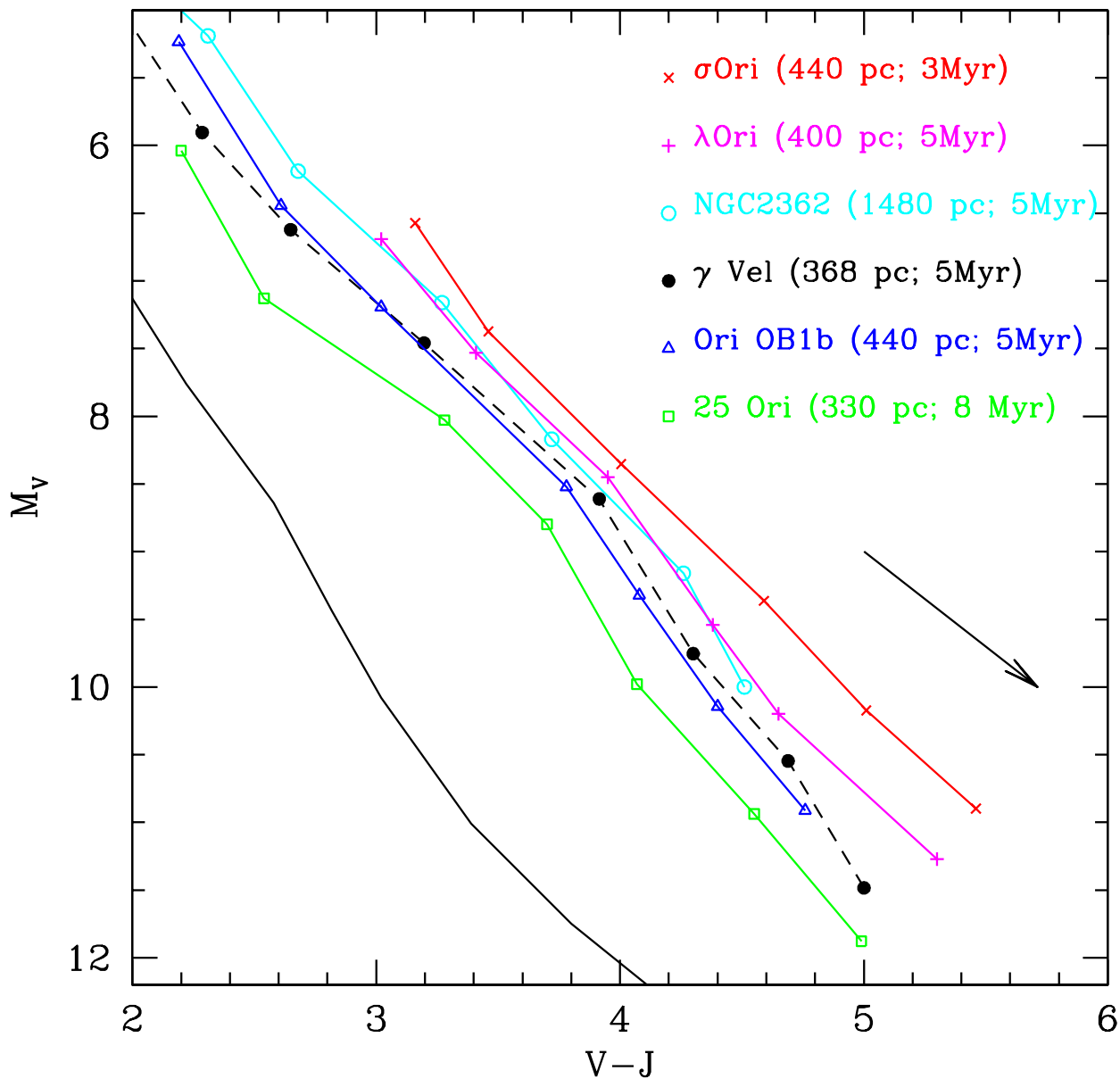}
\caption{Color - absolute magnitude diagram comparing 
the median color and magnitudes of members of several 
5 Myr old stellar populations and the $\gamma$ Velorum cluster. As reference,
the $\sim$8Myr old cluster 25 Orionis and the $\sim$3 Myr 
old cluster $\sigma$ Orionis are also displayed.  The arrow indicates the reddening vector
for \av=1. It is apparent that the 
$\gamma$ Velorum cluster has similar colors to the other 5 Myr old stellar groups.}
\label{fig:cmd3}
\end{figure}

To support the assumed age, we plot in Figure \ref{fig:cmd3} the 
median absolute magnitudes versus the median V-J color for  members 
of several different stellar groups.  
Regardless of the uncertainties introduced by theoretical 
evolutionary tracks, Figure \ref{fig:cmd3} shows 
that the $\gamma$ Velorum cluster is in a similar evolutionary
stage as the other stellar groups with ages normally quoted 
as $\sim$5 Myr: the $\lambda$ Orionis cluster  \citep{barrado07, dolan02}, 
the cluster NGC 2362 \citep{dahm07}, and the Orion OB1b subassociation 
\citep{briceno07,hernandez07b}. We estimate an error of 1.5 Myr comparing 
the standard deviation  in the color V-J for the member sample 
(see \S\ref{s:mem}) to the standard deviation obtained 
using the theoretical isochrones from \citet{sf00} with a reference age of 5 Myr.

\begin{figure}
\epsscale{1.0}
\plotone{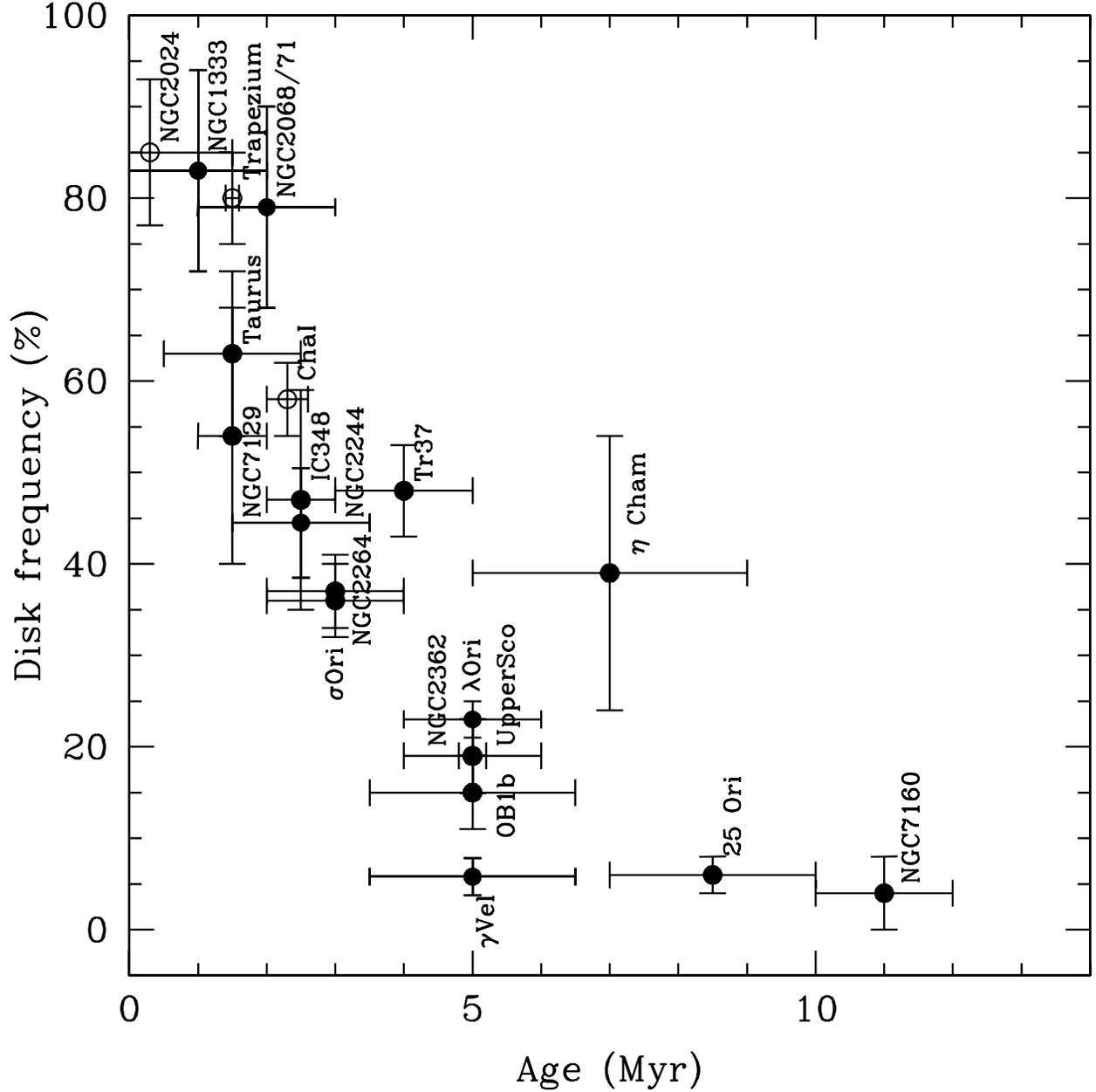}
\caption{ Fraction of stars with near-infrared disk emission as a function
of the age of the stellar group. Open circles represent the disk frequency 
for stars in the T Tauri mass range (TTS; $\sim$K5 or later), derived using JHKL observations:
NGC2024 and Trapezium \citep{haisch01},  and Chameleon I \citep{gomez01}.
Solid symbols represent the disk frequency calculated for stars in the TTS mass 
range using Spitzer data (see text for references).}
\label{fig:fracD}
\end{figure}

Figure \ref{fig:fracD} shows the disk frequencies of late type stars 
(stars K middle or later) with near-infrared disk emission in different stellar 
groups, as a function of age \citep{hernandez05,hernandez07a,haisch01}.
Using the number of members with V-J$>$2.0 ($\sim$K5 or later) 
expected  in our photometric sample (dashed histogram in 
Figure \ref{fig:cmd2}) and the disk detected in  \S \ref{s:late}, 
we calculated a primordial disk frequency of 6$\pm$2\% for 
the $\gamma$ Velorum cluster.
We include recent Spitzer results for the young stellar clusters 
NGC 1333 \citep{gutermuth07}, NGC 2068/71\citep{flaherty07}, Taurus \citep{hartmann05a},
NGC 7129 \citep{gutermuth04}, Chameleons \citep{megeath05}, Tr 37 and NGC 7160
\citep{aurora06}, IC 348 \citep{lada06}, NGC 2244 \citep{balog07}, 
NGC 2264 \citep{cieza07}, $\sigma$ Ori \citep{hernandez07a}, NGC 2362 \citep{dahm07}, $\lambda$ Ori \citep{barrado07},
Upper Scorpius \citep{carpenter06}, and Orion OB1b and 25 Ori \citep{hernandez07b}.

The disk frequencies decrease toward older ages with 
a time scale for primordial disk dissipation of $\sim$5 Myr. 
It is apparent that the disk frequency found in the $\gamma$ Velorum cluster 
is lower than those found in young stellar populations with similar ages, and 
comparable to the disk frequency in older stellar  groups. 
This could indicate that the low disk presence observed in 
the $\gamma$ Velorum cluster is abnormal for its evolutionary stage, 
and environmental effects, such as strong stellar winds and/or strong  radiation 
fields from the $\gamma$ Velorum system, could provide the physical
mechanism for the low disk frequency.
Moreover, $\sim$75\% of the disk bearing stars in the $\gamma$
Velorum cluster show  IRAC SED slopes smaller than the median values
of other 5 Myr old stellar groups plotted in  Figure \ref{fig:cmd3}
suggesting that the disks of the $\gamma$ Velorum cluster 
have a higher degree of dust settling. In particular, the median 
IRAC SED slopes for the disk population of NGC2362, 
the $\lambda$ Orionis cluster, the OB1b subassociation 
and the $\gamma$ Velorum cluster are 
-1.72, -1.60, -1.70 and -1.82, respectively (with a typical error of 0.06).

Studying the young (2-3 Myr) open cluster NGC2244,  \citet{balog07} showed that 
high-mass stars (O type stars) can affect the primordial disks of
lower mass members only if they are  
within $\sim$0.5 pc of the high-mass star.
We find similar results for the $\gamma$ Velorum cluster. 
Using the photometric members with V-J$>$3.5 ($\sim$M2 or later),
we find a disk frequency of 4$\pm$3\% at projected distance of 
0.25-1.0 pc; inside 0.25 pc the central objects of the cluster contaminate 
the optical photometry used to select photometric candidates (\S \ref{s:mem}).
The closest primordial disk is located at projected distance of  $\sim$0.5 pc.
At larger projected distance ($>$1.0 pc), the disk  frequency is larger (8$\pm$2\%).  
This suggests that the relative fast dispersion of disks in the $\gamma$ 
Velorum cluster  is produced by the strong radiation fields 
and strong stellar winds from the central objects.  
This result must still be considered tentative given 
the small number of stars with disks,  which results 
in large errors in disk frequencies.

Alternatively, Figure \ref{fig:fracD} indicates 
that the disk frequency drops rapidly at $\sim$5 Myr 
and the relative low frequency of disks 
observed in the $\gamma$ Velorum cluster 
could be explained if the cluster is slightly 
older than 5 Myr. Additional studies of photometric 
members presented in Table \ref{tab:members} are necessary 
to explain the disk population found in 
 $\gamma$ Velorum cluster.

\section{Conclusions}
\label{sec:conc}
We have used the IRAC and MIPS instruments on board the Spitzer Space Telescope
to conduct a study of disks around the $\gamma$ Velorum cluster. Since the 
central object is a binary system consisting of the closest known Wolf Rayet 
star and a high mass O star, a strong UV radiation field and stellar winds 
are present in the cluster. Using optical photometry of  X ray sources (2XMM) 
and members confirmed by spectroscopy (Jeffries R.D. et al, in preparation), 
579 photometric candidates were selected as possible members of the cluster. 
The level of contamination by non-members depends on the V-J color range, 
showing the highest level of contamination ($\sim$68\%) at V-J = 1.5 - 3.5, 
where the field giant
branch crosses the young stellar population. Combining optical, 2MASS and Spitzer
data we have detected infrared excess in 29 stars. One of the infrared 
excess stars is a Be star. We report 5 debris 
disks around A type stars, 5 debris disks around solar type stars 
(spectral type range F - early-K) and one solar type stars with 
infrared excess produced by a very massive debris disks or by a 
primordial disk with a high degree of dust settling. Seventeen 
disk bearing low mass stars (K5 or later) were found in the cluster 
with a range of disk properties.  We classified these objects in
three classes using the infrared excess at 5.8{\micron} and the 
SED slope [8.0]-[24]:  9 class II stars, 7 evolved disks star 
and one pretransitional candidate. We found that 76\% of the stars bearing
primordial disks have color V-J$>$4 ($\sim$later than M3.5), indicating a mass
dependent timescale for disk dissipation in the $\gamma$ Velorum cluster, similar 
to results in other young star populations 
\citep{carpenter06,hernandez07a,hernandez07b,lada06}.

We found that the disk frequency observed in the $\gamma$ Velorum cluster 
is lower than in other young stellar groups with similar ages ($\sim$5 Myr), 
and comparable to values found in older populations ($\sim$10Myr). 
Empirical isochrones support the age of $\sim$5 Myr for the overall
population of the $\gamma$ Velorum cluster. The relative low frequency 
observed in the cluster is consistent with: (1) rapid disk dissipation due to the influence of
the strong radiation fields and winds from the WR and O star
cluster members or (2) the age of the cluster is slightly older 
and the low frequency reflects the fast dropping in disk frequency 
observed at $\sim$5 Myr.  More than 75\% of the 
primordial disk stars have inner disk emission lower than 
the median observed in the disk population of other 5 Myr clusters, 
indicating that  the disks of the $\gamma$ Velorum cluster 
have a larger degree of dust settling.

We find no primordial disks closer than  $\sim$0.75 pc from the 
high mass central binary.   At projected distance $>$1 pc, 
the disk fraction seems to be constant around 5-7\%. This result is similar 
to that found in the NGC 2244 cluster \citep{balog07} in which primordial 
disks are affected only in the immediate vicinity of the O stars ($<$0.5pc). 
Our results reinforce an emerging trend in studies of disk frequency,
specifically that optically thick disk dissipation is rapid for $\sim 90$\% of
systems, lasting less than 5 Myr, while $\sim 10$\% of disks can last for
8 Myr or more.

\acknowledgements

We thank to Aletta Tibbetts R.D.  for helping us in the initial steps of the IRAC reduction data. 
An anonymous referee provided many insightful comments.
This publication makes use of data products from
Two Micron All Sky Survey, which is a joint project of the University
of Massachusetts and the Infrared Processing and Analysis Center/California
Institute of Technology. This work is based on observations
made with the {\em Spitzer Space Telescope} (GO-1 0037), which is
operated by the Jet Propulsion Laboratory, California Institute of Technology under
a contract with NASA. Support for this work was provided by University of Michigan
and NASA Grants NAG5-13210, NASA-1277575, and NASA-1285169.

\clearpage

\begin{deluxetable}{lllllllllllll}
\rotate
\tabletypesize{\tiny}
\tablewidth{0pt}
\tablecaption{Members of the $\gamma$ Velorum Cluster \label{tab:members}}
\tablehead{
\colhead{ID} & \colhead{2MASS name} & \colhead{B} & \colhead{V} & \colhead{Ic} & \colhead{Ref} & \colhead{[3.6]} & \colhead{[4.5]} & \colhead{[5.8]} & \colhead{[8.0]} & \colhead{Flux$_{24}$} & \colhead{X-ray} & \colhead{Disk} \\
\colhead{  } & \colhead{ }          &\colhead{mag}&\colhead{mag}&\colhead{mag} & \colhead{}    & \colhead{mag}   & \colhead{mag}   & \colhead{mag}   & \colhead{mag}   & \colhead{mJy}        & \colhead{source} & \colhead{type} \\
}
\startdata
1 & 08064389-4731532 & 19.21 $\pm$ 0.02 & 17.65 $\pm$ 0.01 & 14.98 $\pm$ 0.01 & 1 & 12.39 $\pm$ 0.03 & 12.38 $\pm$ 0.03 & 12.33 $\pm$ 0.04 & 12.32 $\pm$ 0.04 & \nodata & 0 & CIII \\
2 & 08064805-4728118 & 13.43 $\pm$ 0.01 & 12.31 $\pm$ 0.01 & 11.15 $\pm$ 0.01 & 1 & 9.52 $\pm$ 0.03 & 9.57 $\pm$ 0.03 & 9.56 $\pm$ 0.03 & 9.52 $\pm$ 0.03 & \nodata & 0 & CIII \\
3 & 08065006-4732219 & 19.65 $\pm$ 0.03 & 18.22 $\pm$ 0.01 & 15.50 $\pm$ 0.03 & 1 & 12.75 $\pm$ 0.03 & 12.66 $\pm$ 0.03 & 12.66 $\pm$ 0.06 & 12.71 $\pm$ 0.07 & \nodata & 0 & CIII \\
4 & 08065158-4728058 & 14.55 $\pm$ 0.01 & 13.34 $\pm$ 0.01 & 12.03 $\pm$ 0.01 & 1 & 10.20 $\pm$ 0.03 & 10.25 $\pm$ 0.03 & 10.21 $\pm$ 0.03 & 10.19 $\pm$ 0.03 & \nodata & 0 & CIII \\
5 & 08065419-4730210 & 14.54 $\pm$ 0.01 & 13.34 $\pm$ 0.01 & 12.05 $\pm$ 0.01 & 1 & 10.23 $\pm$ 0.03 & 10.25 $\pm$ 0.03 & 10.20 $\pm$ 0.03 & 10.17 $\pm$ 0.03 & \nodata & 0 & CIII \\
6 & 08070052-4732093 & 13.58 $\pm$ 0.01 & 12.82 $\pm$ 0.01 & 12.03 $\pm$ 0.01 & 1 & 10.98 $\pm$ 0.03 & 11.01 $\pm$ 0.03 & 10.99 $\pm$ 0.03 & 11.00 $\pm$ 0.03 & \nodata & 0 & CIII \\
7 & 08070716-4721462 & 12.99 $\pm$ 0.01 & 11.90 $\pm$ 0.01 & 10.74 $\pm$ 0.01 & 1 & 9.16 $\pm$ 0.03 & 9.18 $\pm$ 0.03 & 9.16 $\pm$ 0.03 & 9.12 $\pm$ 0.03 & 1.59 $\pm$ 0.61 & 0 & CIII \\
8 & 08071155-4719512 & 16.09 $\pm$ 0.01 & 14.77 $\pm$ 0.01 & 13.13 $\pm$ 0.01 & 1 & 11.10 $\pm$ 0.03 & 11.10 $\pm$ 0.03 & 11.07 $\pm$ 0.03 & 11.05 $\pm$ 0.03 & \nodata & 0 & CIII \\
9 & 08071312-4721124 & 18.05 $\pm$ 0.01 & 16.60 $\pm$ 0.01 & 14.03 $\pm$ 0.01 & 1 & 11.35 $\pm$ 0.03 & 11.33 $\pm$ 0.03 & 11.26 $\pm$ 0.03 & 11.22 $\pm$ 0.03 & \nodata & 0 & CIII \\
10 & 08071363-4725155 & 13.39 $\pm$ 0.01 & 12.73 $\pm$ 0.01 & 11.94 $\pm$ 0.01 & 1 & 10.86 $\pm$ 0.03 & 10.84 $\pm$ 0.03 & 10.81 $\pm$ 0.03 & 10.83 $\pm$ 0.03 & \nodata & 0 & CIII \\

\enddata

\tablenotetext{~}{References in columns 6: 1 this work;  2 \citet{kharchenko01}}
\tablenotetext{~}{Column 12: 1 X-ray emission; 0 not X-ray emission}
\tablenotetext{~}{Column 13: CIII=diskless, CII=optically thick disk, EV= evolved disk, PTD=pretransition disk candidate, DD=Debris disk candidate, Mgiant: Giant candidate based on 2MASS colors}
\tablecomments{Table \ref{tab:members} is published in its entirety in the electronic edition of the {\it Astrophysical Journal}. Electronic edition 
includes 2MASS photometry}.
\end{deluxetable}

\clearpage

\begin{thebibliography}

\bibitem[Abt et al.(1976)]{abt76} 
Abt, H.~A., Landolt, A.~U., Levy, S.~G., \& Mochnacki, S.\ 1976, \aj, 81, 541 


\bibitem[Adams et al.(2004)]{adams04} 
Adams, F.~C., Hollenbach, D., Laughlin, G., \& Gorti, U.\ 2004, \apj, 611, 360 

\bibitem[Balog et al.(2007)]{balog07} Balog, Z., Muzerolle, J., 
Rieke, G.~H., Su, K.~Y.~L., Young, E.~T., \& Megeath, S.~T.\ 2007, \apj, 
660, 1532 

\bibitem[Baraffe et al.(1998)]{baraffe98} 
Baraffe, I., Chabrier, G., Allard, F., \& Hauschildt, P.~H.\ 1998, \aap, 337, 403 

\bibitem[Barrado y Navascu{\'e}s et al.(2007)]{barrado07} 
Barrado y Navascu{\'e}s, D., et al.\ 2007, \apj, 664, 481 

\bibitem[B{\'e}jar et al.(1999)]{bejar99}
B{\'e}jar, V.~J.~S., Zapatero Osorio, M.~R. \& Rebolo, R.\ 1999, \apj, 521, 671

\bibitem[Bessell \& Brett(1988)]{bessell88} 
Bessell, M.~S., \& Brett, J.~M.\ 1988, \pasp, 100, 1134

\bibitem[Brice{\~n}o et al.(2007)]{briceno07} 
Brice{\~n}o, C., Hartmann, L., Hern{\'a}ndez, J., Calvet, N., Vivas, A.~K., Furesz, G., \& 
Szentgyorgyi, A.\ 2007, \apj, 661, 1119 

\bibitem[Brown et al.(2008)]{brown08} Brown, J.~M., Blake, 
G.~A., Qi, C., Dullemond, C.~P., \& Wilner, D.~J.\ 2008, \apjl, 675, L109 

\bibitem[Burningham et al.(2006)]{burningham06} 
Burningham, B., 2005, PhD thesis, University of Exeter. 

\bibitem[Calvet et al.(2002)]{calvet02}
Calvet, N., D'Alessio, P., Hartmann, L., Wilner, D., Walsh, A., \& Sitko, M.\ 2002, \apj, 568, 1008

\bibitem[Calvet et al.(2005)]{calvet05}
Calvet, N., et al.\ 2005, \apjl, 630, L185

\bibitem[Cardelli et al.(1989)]{cardelli89} 
Cardelli, J.~A., Clayton, G.~C., \& Mathis, J.~S.\ 1989, \apj, 345, 245 

\bibitem[Clarke(2007)]{clarke07} 
Clarke, C.~J.\ 2007, \mnras, 376, 1350 


\bibitem[Carpenter et al.(2006)]{carpenter06} Carpenter, J.~M., 
Mamajek, E.~E., Hillenbrand, L.~A., \& Meyer, M.~R.\ 2006, \apj, 651, 49 

\bibitem[Chen et al.(2005)]{chen05} Chen, C.~H., Jura, M., 
Gordon, K.~D., \& Blaylock, M.\ 2005, \apj, 623, 493 


\bibitem[Cieza \& Baliber(2007)]{cieza07} Cieza, L., \& 
Baliber, N.\ 2007, \apj, 671, 605 


\bibitem[Crowther(2007)]{crowther07} 
Crowther, P.~A.\ 2007, \araa, 45, 177

\bibitem[Currie et al.(2008)]{currie08} 
Currie, T., Kenyon, S.~J., Balog, Z., Rieke, G., Bragg, A., \& Bromley, B.\ 2008, \apj, 672, 
558 

\bibitem[Cutri et al.(2003)]{cutri03}
Cutri, R.~M., et al.\ 2003, VizieR Online Data Catalog, 2246, 0.


\bibitem[Dahm \& Hillenbrand(2007)]{dahm07} 
Dahm, S.~E., \& Hillenbrand, L.~A.\ 2007, \aj, 133, 2072 

\bibitem[D'Alessio et al.(2005a)]{dalessio05a}
D'Alessio, P., Mer{\'{\i}}n, B., Calvet, N., Hartmann, L., \& Montesinos, B.\ 2005a,
Revista Mexicana de Astronomia y Astrofisica, 41, 61.

\bibitem[D'Alessio et al.(2005b)]{dalessio05b}
D'Alessio, P., et al.\ 2005b, \apj, 621, 461

\bibitem[D'Alessio et al.(2006)]{dalessio06} D'Alessio, P., 
Calvet, N., Hartmann, L., Franco-Hern{\'a}ndez, R., \& Serv{\'{\i}}n, H.\ 
2006, \apj, 638, 314

\bibitem[De Marco et al.(2000)]{demarco00} De Marco, O., Schmutz, 
W., Crowther, P.~A., Hillier, D.~J., Dessart, L., de Koter, A., \& 
Schweickhardt, J.\ 2000, \aap, 358, 187 


\bibitem[De Marco \& Schmutz(1999)]{demarco99} De Marco, O., \& 
Schmutz, W.\ 1999, \aap, 345, 163 


\bibitem[de Zeeuw et al.(1999)]{dezeeuw99} de Zeeuw, P.~T., 
Hoogerwerf, R., de Bruijne, J.~H.~J., Brown, A.~G.~A., \& Blaauw, A.\ 1999, 
\aj, 117, 354 

\bibitem[Dolan \& Mathieu(2002)]{dolan02} 
Dolan, C.~J., \& Mathieu, R.~D.\ 2002, \aj, 123, 387 

\bibitem[Eggen(1980)]{eggen80} 
Eggen, O.~J.\ 1980, \apj, 238, 627 

\bibitem[Eggen(1983)]{eggen83} 
Eggen, O.~J.\ 1983, \aj, 88, 197 

\bibitem[Eggen(1986)]{eggen86} 
Eggen, O.~J.\ 1986, \aj, 92, 1074 

\bibitem[Engelbracht et al.(2007)]{engelbracht07} 
Engelbracht, C.~W., et al.\ 2007, \pasp, 119, 994

\bibitem[Espaillat et al.(2007a)]{espaillat07a} Espaillat, C., et 
al.\ 2007a, \apjl, 664, L111

\bibitem[Espaillat et al.(2007b)]{espaillat07b} Espaillat, C., 
Calvet, N., D'Alessio, P., Hern{\'a}ndez, J., Qi, C., Hartmann, L., Furlan, 
E., \& Watson, D.~M.\ 2007b, \apjl, 670, L135 

\bibitem[Fazio et al.(2004)]{fazio04} 
Fazio, G.~G., et al.\ 2004, \apjs, 154, 39 

\bibitem[Flaherty \& Muzerolle(2007)]{flaherty07} Flaherty, K.~M., 
\& Muzerolle, J.\ 2007, ArXiv e-prints, 712, arXiv:0712.1601

\bibitem[Furlan et al.(2006)]{furlan06} 
Furlan, E., et al.\ 2006, \apjs, 165, 568 

\bibitem[G{\'o}mez \& Kenyon(2001)]{gomez01} 
G{\'o}mez, M., \& Kenyon, S.~J.\ 2001, \aj, 121, 974 

\bibitem[Gordon et al.(2005)]{gordon05}
Gordon, K.~D., et al.\ 2005, \pasp, 117, 503

\bibitem[Gorlova et al.(2004)]{gorlova04} 
Gorlova, N., et al.\ 2004, \apjs, 154, 448 

\bibitem[Gorlova et al.(2006)]{gorlova06} Gorlova, N., Rieke, 
G.~H., Muzerolle, J., Stauffer, J.~R., Siegler, N., Young, E.~T., \& 
Stansberry, J.~H.\ 2006, \apj, 649, 1028 

\bibitem[Gorlova et al.(2007)]{gorlova07} 
Gorlova, N., Balog, Z., Rieke, G.~H., Muzerolle, J., Su, K.~Y.~L., Ivanov, V.~D., \& Young, E.~T.\ 
2007, \apj, 670, 516 

\bibitem[Gutermuth et al.(2004)]{gutermuth04}
Gutermuth, R.~A., Megeath, S.~T., Muzerolle, J., Allen, L.~E., Pipher, J.~L., Myers, P.~C.,
\& Fazio, G.~G.\ 2004, \apjs, 154, 374

\bibitem[Gutermuth et al.(2007)]{gutermuth07} Gutermuth, R.~A., et 
al.\ 2007, ArXiv e-prints, 710, arXiv:0710.1860 

\bibitem[Haisch et al.(2001)]{haisch01} 
Haisch, K.~E., Lada, E.~A., \& Lada, C.~J.\ 2001, \apjl, 553, L153

\bibitem[Hartmann(2003)]{hartmann03} Hartmann, L.\ 2003, \apj, 
585, 398 

\bibitem[Hartmann et al.(2005)]{hartmann05a} 
Hartmann, L., Megeath, S.~T., Allen, L., Luhman, K., Calvet, N., D'Alessio, P., Franco-Hernandez,
R., \& Fazio, G.\ 2005, \apj, 629, 881

\bibitem[Hartmann(2005)]{hartmann05b} 
Hartmann, L.\ 2005, ASP Conf.~Ser.~341: Chondrites and the Protoplanetary Disk, 341, 131 


\bibitem[Hern{\'a}ndez et al.(2005)]{hernandez05} 
Hern{\'a}ndez, J., Calvet, N., Hartmann, L., Brice{\~n}o, C., Sicilia-Aguilar, A., \& Berlind, P.\ 2005, \aj, 129, 856.

\bibitem[Hern{\'a}ndez et al.(2006)]{hernandez06} Hern{\'a}ndez, 
J., Brice{\~n}o, C., Calvet, N., Hartmann, L., Muzerolle, J., \& Quintero, 
A.\ 2006, \apj, 652, 472 

\bibitem[Hern{\'a}ndez et al.(2007a)]{hernandez07a} Hern{\'a}ndez, 
J., et al.\ 2007a, \apj, 662, 1067

\bibitem[Hern{\'a}ndez et al.(2007b)]{hernandez07b} Hern{\'a}ndez, J., et 
al.\ 2007b, \apj, 671, 1784

\bibitem[Hollenbach \& Adams(2004)]{hollenbach04} 
Hollenbach, D., \& Adams, F.~C.\ 2004, Star Formation in the Interstellar Medium: In Honor of 
David Hollenbach, 323, 3 

\bibitem[Houk(1978)]{houk78}
Houk, N., 1978, Michigan Catalogue for the HD Stars, Vol. 2 
(Ann Arbor: Dept. Astron., Univ. Michigan), 
VizieR Online Data Catalog III/51B

\bibitem[Jeffries et al.(2000)]{jeffries00} 
Jeffries, R.D., Pozzo, M., Naylor, T., Harmer, S., Walter, F.M. 2000 in
"X-ray astronomy 2000" eds. R. Giacconi, S. Seria, L. Stella, ASP Conference Series
Vol. 234

\bibitem[Kenyon \& Bromley(2005)]{kenyon05}
Kenyon, S.~J.~\& Bromley, B.\ 2005, \aj, 130, 269

\bibitem[Kenyon \& Hartmann(1995)]{kh95}
Kenyon, S.~J.~\& Hartmann, L.\ 1995, \apjs, 101, 117

\bibitem[Kharchenko(2001)]{kharchenko01} 
Kharchenko, N.~V.\ 2001, Kinematika i Fizika Nebesnykh Tel, 17, 409.

\bibitem[Lada et al.(2006)]{lada06} 
Lada, C.~J., et al.\ 2006, \aj, 131, 1574 

\bibitem[Landsman(1993)]{landsman93} 
Landsman, W.~B.\ 1993, ASP Conf.~Ser.~ 52: Astronomical Data Analysis Software and Systems II, 52, 246 

\bibitem[Lyra et al.(2006)]{lyra06} Lyra, W., Moitinho, A., 
van der Bliek, N.~S., \& Alves, J.\ 2006, \aap, 453, 101 

\bibitem[MacConnell(1981)]{macconnell81} 
MacConnell, D.~J.\ 1981, \aaps, 44, 387

\bibitem[Ma{\'{\i}}z Apell{\'a}niz et al.(2008)]{maiz08} 
Ma{\'{\i}}z Apell{\'a}niz, J., Alfaro, E.~J., 
\& Sota, A.\ 2008, ArXiv e-prints, 804, arXiv:0804.2553 

\bibitem[Megeath et al.(2005a)]{megeath05} 
Megeath, S.~T., Hartmann, L., Luhman, K.~L., \& Fazio, G.~G.\ 2005a, \apjl, 634, L113 

\bibitem[Megeath et al.(2007)]{megeath07} Megeath, S.~T., Gaidos, 
E., Hester, J.~J., Adams, F.~C., Bally, J., Lee, J.~-., \& Wolk, S.\ 2007, 
ArXiv e-prints, 704, arXiv:0704.1045 


\bibitem[Meyer et al.(2008)]{meyer08} 
Meyer, M.~R., et al.\ 2008, \apjl, 673, L181 


\bibitem[Meyer et al.(1997)]{meyer97} 
Meyer, M.~R., Calvet,  N., \& Hillenbrand, N. A.\ 1997, \aj, 114, 288 

\bibitem[Millour et al.(2007)]{millour07} Millour, F., et al.\ 
2007, \aap, 464, 107 


\bibitem[Naylor(1998)]{naylor98} Naylor, T.\ 1998, \mnras, 296, 
339 

\bibitem[Naylor et al.(2002)]{naylor02} Naylor, T., Totten, 
E.~J., Jeffries, R.~D., Pozzo, M., Devey, C.~R., \& Thompson, S.~A.\ 2002, 
\mnras, 335, 291

\bibitem[North et al.(2007)]{north07} North, J.~R., Tuthill, 
P.~G., Tango, W.~J., \& Davis, J.\ 2007, \mnras, 377, 415 


\bibitem[Pozzo et al.(2000)]{pozzo00} 
Pozzo, M., Jeffries, R.~D., Naylor, T., Totten, E.~J., Harmer, S., \& Kenyon, M.\ 2000, \mnras, 
313, L23 

\bibitem[Reach et al. (2006)]{reach06} 
Reach, W. et al. \ 2006, Infrared Array Camera Data Handbook, version 3.0, 
{\em Spitzer} Science Center, California Institute of Technology, Pasadena, California 91125 USA. 

\bibitem[Richling \& Yorke(2000)]{richling00} Richling, S., \& 
Yorke, H.~W.\ 2000, \apj, 539, 258

\bibitem[Rieke et al.(2004)]{rieke04}
Rieke, G. H., et al, 2004, \apjs, 154, 25

\bibitem[Rieke et al.(2005)]{rieke05}
Rieke, G. H., et al, 2005, \apj, 620, 1010

\bibitem[Schaerer et al.(1997)]{schaerer97} 
Schaerer, D., Schmutz, W., \& Grenon, M.\ 1997, \apjl, 484, L153

\bibitem[Sicilia-Aguilar et al.(2006)]{aurora06}
Sicilia-Aguilar, A., et al.\ 2006, \apj, 638, 897

\bibitem[Siegler et al.(2007)]{siegler07} Siegler, N., Muzerolle, 
J., Young, E.~T., Rieke, G.~H., Mamajek, E.~E., Trilling, D.~E., Gorlova, 
N., \& Su, K.~Y.~L.\ 2007, \apj, 654, 580 


\bibitem[Siess et al.(2000)]{sf00}
Siess, L., Dufour, E., \& Forestini, M.\ 2000, \aap, 358, 593

\bibitem[Strom et al.(1989)]{strom89} 
Strom, K.~M., Strom, S.~E., Edwards, S., Cabrit, S., \& Skrutskie, M.~F.\ 1989, \aj, 97, 1451 

\bibitem[Throop \& Bally(2005)]{throop05} Throop, H.~B., \& 
Bally, J.\ 2005, \apjl, 623, L149

\bibitem[van Leeuwen(2007)]{leeuwen07} 
van Leeuwen, F.\ 2007, \aap, 474, 653 

\end{thebibliography}
\end{document}